\documentclass[prd,aps,showpacs,nofootinbib,eqsecnum,onecolumn,groupedaddress,amssymb]{revtex4}
\usepackage{amsmath}
\usepackage{amssymb}
\usepackage{amsfonts}
\usepackage{graphicx,bm}
\usepackage{dcolumn}
\usepackage[colorlinks=true]{hyperref}
\usepackage{color,amsxtra}
\usepackage{epsf}
\usepackage{enumerate}
\usepackage{hhline}
\usepackage{array}
\usepackage{tabularx}
\usepackage{subfigure}
\usepackage{fancyhdr}
\usepackage{mathrsfs}

\newcommand{\be}{\begin{equation}}
\newcommand{\ee}{\end{equation}}
\newcommand{\bea}{\begin{eqnarray}}
\newcommand{\eea}{\end{eqnarray}}
\newcommand{\beaa}{\begin{eqnarray*}}
\newcommand{\eeaa}{\end{eqnarray*}}

\newcommand{\nn}{\nonumber \\}
\newcommand{\e}{\mathrm{e}}

\def\be{\begin{equation}}
\def\ee{\end{equation}}
\def\bea{\begin{eqnarray}}
\def\eea{\end{eqnarray}}

\begin{document}


\title{Testing logarithmic corrections on $R^2$-exponential gravity by observational data}


\author{Sergei D. Odintsov}
\email{odintsov@ice.csic.es} \affiliation{Institut de Ci\`{e}ncies de l'Espai,
ICE/CSIC-IEEC, Campus UAB, Carrer de Can Magrans s/n, 08193 Bellaterra (Barcelona),
Spain} \affiliation{Instituci\'o Catalana de Recerca i Estudis Avan\c{c}ats (ICREA),
Barcelona, Spain}
\author{Diego S\'aez-Chill\'on G\'omez}
\email{diego.saez@ehu.eus} \affiliation{Department of Theoretical Physics, University of the Basque Country UPV/EHU, \\ P.O. Box 644, 48080 Bilbao, Spain}
\author{German~S.~Sharov}
\email{sharov.gs@tversu.ru} \affiliation{Tver state university 170002, Sadovyj per. 35,
Tver, Russia}

%

%
%
\begin{abstract}

This paper is devoted to the analysis of a class of $F(R)$ gravity, where additional
logarithmic corrections are assumed. The gravitational action includes an exponential
term and a $R^2$ inflationary term, both with logarithmic corrections. This model can
unify an early time inflationary era and also the late time acceleration of the universe
expansion. This model is deeply analysed, confronting with recent observational data
coming from
 the largest Pantheon Type Ia supernovae sample,
 the latest measurements of the
Hubble parameter $H(z)$, manifestations of Baryon Acoustic Oscillations and Cosmic
Microwave Background radiation. The viability of the model is studied and the
corresponding constraints on the free parameters are obtained, leading to an statistical
analysis in comparison to  $\Lambda$CDM model. The inflationary era is also analysed
within this model and its compatibility with the latest observational data for the
spectral index of primordial curvature perturbations and the scalar-to-tensor ratio.
Finally, possible corrections on the Newton's law and constraints due to
primordial nucleosynthesis are analysed.
\end{abstract}
\maketitle
%
%
%

\section{Introduction}

Some modifications of General Relativity (GR) have drawn a lot of attention over the last years. Most of them keep the basic principles of GR as the Equivalence Principle and General Covariance, but focus on modifications of the field equations, leading to new solutions and in general to more complexity. Besides the inherent academic value of studying extensions of GR for understanding gravity and geometry better, some modifications of GR have been proposed to provide a way for explaining some of the most important challenges in cosmology nowadays, as dark energy, inflation or dark matter (for a review see \cite{reviews1}). Nevertheless, while most of the attempts to explain dark matter through a new gravitational theory have not provided a reliable and successful scenario, modified gravities seem much more promising to explain the conundrum of dark energy and also inflation \cite{Capozziello2002,unifying,Joyce:2014kja}. \\

Particularly, some of the most successful inflationary models are based on generalisations of the Einstein-Hilbert action, the so-called $f(R)$ gravity, mainly due to the ease to reconstruct the appropriate action capable of reproducing an accelerating expansion of the universe, as also occurs during the dark energy epoch. In addition, last data released by Planck collaboration \cite{Planck-Inflation} on the Cosmic Microwave Background (CMB), infer a very small rate of the power spectrum for the gravitational waves background generated during inflation, a constraint that has ruled out some inflationary models, but keeps $f(R)$  gravity, particularly the so-called $R^2$ (Starobinsky) inflation \cite{Starobinsky1980}, as one of the most promising candidates for inflation. Moreover, late-time acceleration is also realised in $f(R)$ gravities, where any particular solution may find its corresponding gravitational action \cite{unifying}. To do so, a number of techniques have been developed in order to deal with a theory that own fourth field order equations, but which can be decomposed by an scalar field, reducing to a type of Brans-Dicke theory \cite{reviews1}. Nevertheless, any modification of GR may introduce severe corrections on well tested results, particularly on local gravity tests. To deal with that inconvenient, an screening mechanism called chameleon mechanism \cite{Khoury:2003rn}, originally proposed to hide light scalar fields at different scales is applied to $f(R)$ gravities leading to some particular gravitational actions that accomplish a number of viability conditions \cite{Pogosian:2007sw} and produce the desirable late-time acceleration of the universe expansion, generally mimicking a cosmological constant at late-times \cite{Hu:2007nk} but also with additional terms that may include the inflationary epoch \cite{Nojiri:2007cq}. These models have drawn a lot of attention, specially due to the strong increase of data, both describing the early stages of the universe as late time epochs, such that  any desirable cosmological scenario, and its corresponding $f(R)$ action in this case, should satisfy observational limitations for both early and late-time acceleration eras, as well as theoretical constraints \cite{Appleby:2007vb}.\\

One of these types of $f(R)$ models is the so-called exponential gravity, which includes an exponential function of the Ricci scalar in the action of the form$e^{-R/R_*}$, such that by the appropriate choice for the constant $R_*$, the scale on which that the exponential plays a role can be easily managed \cite{Linder2009}. This is important along the cosmological history, since the universe goes through different stages, each one characterised by a different value of the Ricci scalar. Since the exponential acts basically as an step function, with a fast transition at $R\sim R_*$, the term acts as an effective cosmological elsewhere, what can be used to mimic $\Lambda$CDM model, as suggested in the literature \cite{Linder2009,nostriexp}, satisfying the observational constraints \cite{YangLeeLG2010,OdintsovSGS:2017}. In addition, the model may be implemented in such a way that includes vacuum solutions as Minkowski or Schwarzschild as solutions, in comparison to the presence of a cosmological constant. Moreover, by the appropriate scale, an effective inflationary phase can be included, leading to a gravitational action that may be capable of reproducing the whole cosmological evolution \cite{nostriexp}. Actually, such exponential may be used to find possible corrections and tests to $R^2$ inflation and to suppress the effects of inflationary terms at later times \cite{OdintsovSGS:2017}. Such type of models has been well tested and compared to other models, leading to very promising results for describing the whole cosmological history.\\

In this paper we consider a particular exponential gravity, where some extra terms are
included in the action in order to test the reliability of exponential models as well as
the $\Lambda$CDM model. Here, the extra terms in the action have the form of logarithmic
functions of the Ricci scalar, since the correction evolutes very smoothly in comparison
to the original model and may provide the correct predictions during inflation, as shown
in
Ref.~\cite{Nojiri:2003ni,Cognola:2005de,Elizalde:2017mrn,Myrzakulov:2014hca,OdintsovOS:2017log,Liu:2018hno}.
Such type of logarithmic corrections are induced by quantum gravity effects, such that
its analysis becomes essential to understand well their behaviour
\cite{Cognola:2005de,Elizalde:2017mrn,Buchbinder:1992rb}. Here a complete gravitational
Lagrangian is provided, composed by some exponential terms responsible of the dark
energy epoch and corrected by an extra logarithmic, while an $R^2$ term drives the
inflationary epoch but modelled by another logarithmic of the Ricci scalar. Then, we
study in detail how this type of models describes the recent observational data, in
particular, we use the latest Pantheon Type Ia supernovae sample (SNe Ia)
in comparison with  Union 2.1 SNe Ia observations,
  estimations of the Hubble parameter
$H(z)$, data from baryon acoustic oscillations (BAO) and from cosmic microwave
background radiation (CMB). We calculate the best fit for the free parameters of the
model and compare this model with its analog exponential without logarithmic corrections
and with the standard $\Lambda$CDM model \cite{OdintsovSGS:2017}. Finally, we also
consider in detail the inflationary epoch, its observable manifestations and the
viability of the full Lagrangian during the whole
cosmological evolution of the Universe. \\

The paper is organised as follows: In section \ref{FRsection}, we briefly review $f(R)$
gravity, its corresponding equations and the Lagrangian on which the paper is based.
Section \ref{Late} is devoted to the analysis of the model along the cosmological
evolution after inflation. In \ref{Data}, we describe the observational data for SNe Ia,
$H(z)$, BAO and CMB that is used. While in section \ref{Analysis}, we obtain the
constraints on the free parameters of the model and compare to other models. Section
\ref{Inflation} is focused on the inflationary era for model. Section  \ref{Newton} is
devoted to the analysis of possible violations of the Newton's law at local scales. In section \ref{BBN}, we analyse the constraints from primordial (Big Bang)
nucleosynthesis. Finally, section \ref{conclusions} summarises the results of the paper.

\section{$F(R)$ gravity}
\label{FRsection}

Let us start by introducing the basics of $F(R)$ gravity. The general action for $F(R)$ theories is given by:
 $$ 
  S = \frac1{2\kappa^2}\int d^4x \sqrt{-g}\,F(R)  + S^{m}\ .
  $$ 
The field equations are obtained by varying the action with respect to the metric field,
 \begin{equation}
F_R R_{\mu\nu}-\frac F2
g_{\mu\nu}+\big(g_{\mu\nu}g^{\alpha\beta}\nabla_\alpha\nabla_\beta-\nabla_\mu\nabla_\nu\big)F_R
=\kappa^2T_{\mu\nu}\ .
\label{fieldeq}
\end{equation}
Here $F_R\equiv F'(R)$ and  $F_{RR}\equiv F''(R)$. We are interested in the analysis of a particular type of $F(R)$ Lagrangian, which is known to describe both the inflationary epoch as the late-time acceleration \cite{OdintsovOS:2017log}:
 \begin{equation}
 F(R)=R-2\Lambda \big( 1-e^{-bR/\Lambda}\big)
 \left[1-c\frac{R}{4\Lambda}\log\frac{R}{4\Lambda}\right]+
\gamma(R) R^2\ , \label{explog}
\end{equation}
 where $\kappa^2=8\pi G$, $S^{m}$ is the matter Lagrangian, $\Lambda$ is a cosmological constant and the function $\gamma(R)$ accomplishes for deviations with respect to Starobinsky inflation \cite{Starobinsky1980}, being defined as:
 \begin{equation}
\gamma(R)=\gamma_0\left( 1+\gamma_1\log 
 \frac{R}{R_0} 
  \right)\ . \label{gamma}
\end{equation}
The model parameters $b$, $c$, $\gamma_0$, $\gamma_1$, $R_0$ are positive constants, where $R_0$ is the curvature of the Universe at the end of inflation. The second term in (\ref{explog}) is assumed to become important at late times, which differs from usual exponential gravity by the logarithmic term that provides stability to the solutions, as shown in \cite{OdintsovOS:2017log}.  On the other hand, the term $\gamma(R) R^2$ plays an important role during the inflationary epoch, when $R\ge R_0$, and is inspired by one-loop corrections in higher-derivative quantum gravity \cite{Buchbinder:1992rb,Myrzakulov:2014hca}. In addition, such term may provide a graceful exit from inflation, as shown in \cite{OdintsovOS:2017log,Myrzakulov:2014hca}.

Here, we are focusing on the cosmological analysis of the action (\ref{explog}) and how good the model is for reproducing dark energy and inflation. Hence, we assume a spatially-flat Friedmann-Lema\^itre-Robertson-Walker (FLRW) metric
\begin{equation}
\label{FLRW} ds^2 = - dt^2 + a^2(t) \sum_{i=1}^3 \left(dx^i\right)^2\, .
\end{equation}
where $a(t)$ is the scale factor, such that $H=\dot a/a$ is the Hubble parameter and the Ricci scalar reads $R = 6 (2H^2 + \dot H )$, where the dot denotes derivatives with respect to the cosmic time. From the field equations (\ref{fieldeq}), the modified FLRW equations are obtained:
 \begin{eqnarray}
&& \qquad H^2F_R+\frac16(F-RF_R)+H\dot F_R=\frac13 \kappa^2\rho,  \nonumber \\
&& \!\!\!\!\!\! (2\dot H+3H^2)\,F_R+\frac12(F-RF_R)+2H\dot F_R+\ddot F_R=-\kappa^2 p\ .
\label{dsg3}
 \end{eqnarray}
Here we have assumed a perfect fluid for the energy-momentum tensor:
\be
T_{\mu\nu}= (\rho+p)u_{\mu}u_{\nu}+pg_{\mu\nu}\ ,
\ee
where $u_{\mu}u^{\mu}=-1$. Equations (\ref{dsg3}) may be expressed in a more convenient way as \cite{OdintsovSGS:2017}
 \begin{eqnarray}
\frac{dH}{dN}&=&\frac{R}{6H}-2H, \label{dynam1} \\
\frac{dR}{dN}&=&\frac1{F_{RR}}\bigg(\frac{\kappa^2\rho}{3H^2}-F_R+\frac{RF_R-F}{6H^2}\bigg),
 \label{dynam2}\\
 \frac{d\rho}{dN}&=&-3(\rho+p). \label{dynam3} \end{eqnarray}
where  $N=\log a=-\log(1+z)$ is the number of e-folds, with $a(t_0)=1$ fixed at the
present time $t_0$. Eq.~(\ref{dynam3}) is the consequence of the energy conservation
equation $\nabla^\mu T_{\mu\nu}=0$. Solution of the system (\ref{dynam1}) --
(\ref{dynam3}) provides the cosmological evolution for a particular $F(R)$ model and a
particular equation of state $p=p(\rho)$. In the next sections, we analyse the model
(\ref{explog}), which can be split into two parts that do not overlap along the
cosmological history, the dominant during the dark energy epoch and the corresponding
one at inflation.

\section{Late-time acceleration}
\label{Late}

In this section we study the behaviour of the above model at late times, when the Ricci
scalar $R$ is much smaller than the value at the end of inflation $R_0$. In addition, we
can consider the inflationary term $\gamma(R) R^2$ negligible at late times, when $R\sim
4\Lambda<<R_0$ as far as:
 \be \gamma_0\sim R_0^{-2}\ , \quad \text{and} \quad
\gamma_1<<\frac{1}{\text{log}\left(\frac{R_0}{4\Lambda}\right)}\ . \ee
 where $R_0\sim
10^{85}\Lambda$ as calculated in Ref.~\cite{OdintsovOS:2017log}. In order to test the
goodness of the model, we shall use different datasets that include different phases of
the cosmological evolution, as shown below in Sect.~\ref{Data}. In particular, data from
Supernovae Ia \cite{Union21SNe,Pantheon17}, Baryon Acoustic Oscillations (BAO)
\cite{Eisen05}, estimations of the Hubble parameter $H(z)$ corresponding to $z\le2.36$
\cite{Hdata} and parameters of the Cosmic Microwave Background (CMB) from Planck
collaboration \cite{Planck13}. The latter refers to the photon-decoupling epoch at
$z\simeq1100$. Hence, at late times $z<10^5$ we can neglect the inflationary term
$\gamma(R) R^2$ and the $F(R)$ function (\ref{explog}) leads to:
 \begin{equation}
 F(R)=R-2\Lambda \big( 1-e^{-bR/\Lambda}\big)
 \left[1-c\frac{R}{4\Lambda}\log\frac{R}{4\Lambda}\right]=
 2\Lambda\left[ {\cal
R}-\big(1-e^{-\beta{\cal R}}\big)\Big(1-\alpha{\cal R}\log\frac{\cal R}2\Big)\right]\ .
\label{FRlate}
\end{equation}
Here we have redefined the parameters to make them dimensionless as follows
 \begin{equation}
 {\cal R}=\frac{R}{2\Lambda},\qquad \beta=2b,\qquad\alpha=\frac c2
 \label{Rbetaalp}
\end{equation}
In the limit $\alpha=0$ the model (\ref{FRlate}) becomes the usual exponential $F(R)$ model \cite{Linder2009,nostriexp,OdintsovSGS:2017} with no logarithmic corrections, while for $\alpha=0$ together with $\beta\to+\infty$, the model turns out to the standard $\Lambda$CDM Lagrangian $R-2\Lambda$. Note also that the function (\ref{FRlate}) recovers the $\Lambda$CDM model for $\alpha=0$ and $R\to+\infty$. However, for $\alpha\neq 0$ the Lagrangian (\ref{FRlate}) does not recover $\Lambda$CDM model at high redshifts but the logarithmic correction remains:
 \begin{equation}
 F(R)\simeq 2\Lambda\left( {\cal
R}-1+\alpha{\cal R}\log\frac{\cal R}2\right),\qquad\mbox{if}\qquad \beta{\cal R}\gg 1.
\label{FRlimit}\end{equation}
In addition, any $F(R)$ model has to satisfy some particular conditions to be considered as a serious and successful alternative to GR. Hence, in order to keep a positive effective gravitational constant and to avoid the merge of fifth forces, the following condition should hold at high curvature regimes:
\begin{equation}
\left|F_R(R)-1\right|\ll 1 \quad \rightarrow \quad  F_R(R)-1\simeq
\alpha\left(1+\log\frac{R}{4\Lambda}\right)\ll 1\,, \label{alpll1}\end{equation} which
should be satisfied during the post-inflationary era ($4\Lambda\le R<R_0$), particularly
along the radiation, matter dominated eras and late-time acceleration.  Therefore, the
cosmological constant in (\ref{FRlate}) behaves as an attractor at late times, similarly
to $\Lambda$CDM model, as far as the exponential $e^{-\beta{\cal R}}$ remains
negligible. As we are focusing on the post-inflationary period, a pressureless
(non-relativistic) fluid and radiation (relativistic particles) should be included in
the energy-momentum tensor, such that the continuity equation (\ref{dynam3}) can be
solved, leading to:
  \begin{equation}
 \rho=\rho_m^0a^{-3}+ \rho_r^0a^{-4},
 \label{rho}\end{equation}
 where $\rho_m^0$ and $\rho_r^0$ are the energy densities for dust and radiation at the present time, respectively. In order to reduce the number of free parameters, we can fix the radiation-matter ratio as provided by Planck \cite{Planck13}:
\be
X_r=\frac{\rho_r^0}{\rho_m^0}=2.9656\cdot10^{-4}\ .
\label{Xrm}\ee
We shall use dimensionless parameters for the energy densities, which can be expressed in terms of the $\Lambda$CDM model as follows:
 \begin{equation}
  H_0^*\equiv H^{\Lambda CDM}_0, \qquad \Omega_m^*\equiv \Omega_m^{\Lambda CDM}=\frac{\kappa^2\rho_m(t_0)}{3(H_0^*)^2},
 \qquad \Omega_\Lambda^*\equiv \Omega_\Lambda^{\Lambda CDM}=\frac\Lambda{3(H^{*}_0)^2}.
  \label{H0OmOL}\end{equation}
Note that we use  $\Lambda$CDM model as a reference under the assumption that our model will mimic $\Lambda$CDM model far away from the inflationary period. In this sense, the solution of the FLRW equations for the $\Lambda$CDM model reads
 \begin{equation}
 \frac{H^2}{(H^{*}_0)^2}=\Omega_m^{*} \big(a^{-3}+ X_r a^{-4}\big)+\Omega_\Lambda^{*}\ ,\qquad
 \frac{R}{2\Lambda}=2+\frac{\Omega_m^{*}}{2\Omega_\Lambda^{*}}a^{-3}\ .
  \label{asymLCDM}\end{equation}
  Here $X_r\equiv X_r^{\Lambda CDM}$ is the radiation-matter ratio (\ref{Xrm}). In addition, we can redefine the Hubble parameter to have a dimensionless function as follows:
\cite{OdintsovSGS:2017}
\begin{equation}
E=\frac{H}{H_0^{*}}.
   \label{E}\end{equation}
Hence, the dynamical variables  $E(a)$, ${\cal R}(a)$ determine the evolution for the
action (\ref{FRlate}). The corresponding dynamical equations are obtained by assuming
the Lagrangian (\ref{FRlate}) and the density (\ref{rho}) in the equations
(\ref{dynam1}) and (\ref{dynam2}), leading to:
\begin{eqnarray}
\frac{dE}{dN}&=&\Omega_\Lambda^{*}\frac{{\cal R}}{E}-2E, \label{eqE} \\   
\frac{d\log{\cal R}}{dN}&=&\frac{\big\{ E^2_{\Lambda CDM}+\Omega_\Lambda^*\big[\alpha{\cal R}
 \big(1-e^{-\beta {\cal R}}(1-\beta {\cal R}\ell)\big)-e^{-\beta {\cal R}}(1+\beta {\cal R})\big] \big\}\big/E^2-1
 +\beta e^{-\beta {\cal R}} -\alpha\Phi}
 {\alpha+\alpha\,e^{-\beta {\cal R}}\big\{-1+\beta {\cal R}\big[2+(2-\beta {\cal R})\ell \big] \big\}
 +\beta^2 {\cal R}\,e^{-\beta {\cal R}}}.
  \label{eqR}\end{eqnarray}
  Here $\ell=\log({\cal R}/2)$, $\Phi=1+\ell-e^{-\beta {\cal R}}\big[1+(1-\beta {\cal R})\ell \big]$ and
$E^2_{\Lambda CDM}=\Omega_m^{*} \big(a^{-3}+ X_r a^{-4}\big)+\Omega_\Lambda^{*}$, and
recall that the variable $N=\log a$ refers to the number of e-folds. This system can be
solved numerically by setting the appropriate initial conditions. For the model
(\ref{FRlate}) with $\alpha\ne0$ and assuming $\beta{\cal R}\gg 1$, the equation
(\ref{eqR}) in the limit $e^{-\beta{\cal R}}\ll 1$ takes the form
\begin{equation}
\frac{d\log{\cal R}}{dN}\simeq\frac{E^2_{\Lambda CDM}\big/E^2-1}\alpha+
 \Omega_\Lambda^*\frac{\cal R}{E^2}-1-\log\frac{\cal R}2,\qquad \beta{\cal R}\gg 1.
   \label{eqRlim}\end{equation}
This expression accounts for the deviation of our model with respect to the $\Lambda$CDM
model when ${\cal R}$ becomes large enough (${\cal R}\to\infty$ or  $a\to0$).
Nevertheless, the early-time inflation will be considered below in
Sect.~\ref{Inflation}.

\begin{figure}[th]
   \centerline{ \includegraphics[scale=0.7,trim=5mm 0mm -2mm 0mm]{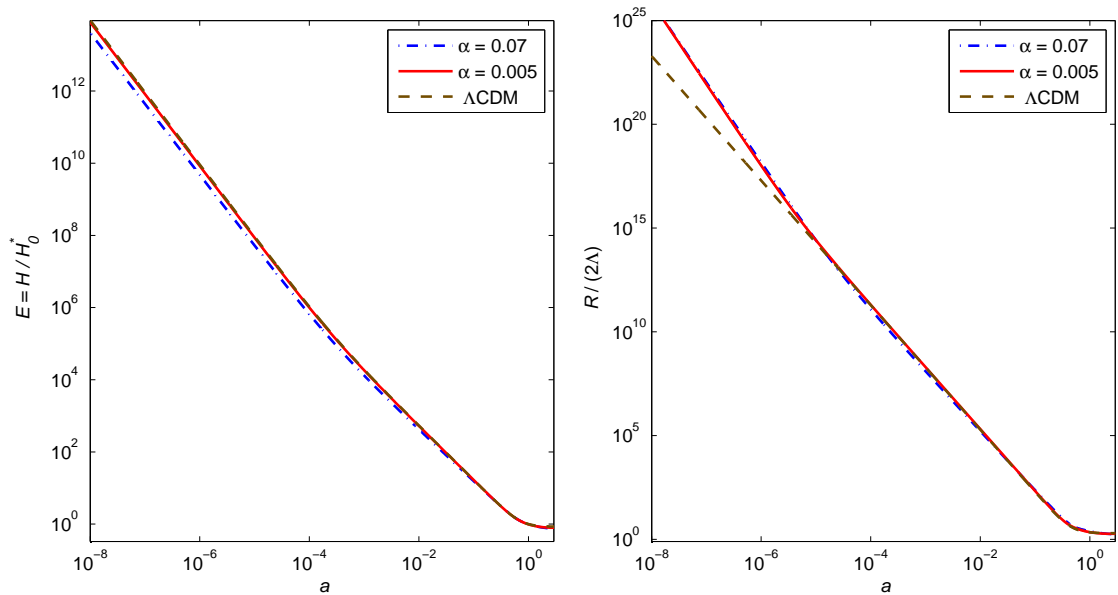}}
\caption{Evolution of the normalized Hubble parameter $E(a)$ and the Ricci scalar ${\cal
R}(a)$ for the $F(R)$ model  (\ref{FRlate}) with 2 sets of the best fitted parameters
(without and with the CMB data) from Table.~\ref{Estim}  (dash-dotted blue lines for
$\alpha=0.07$, $\beta=1.39$, $\Omega_m^*=0.2807$, $\Omega_\Lambda^*=0.587$ and solid red
lines for $\alpha=0.0051$, $\beta=1.95$, $\Omega_m^*=0.2827$, $\Omega_\Lambda^*=0.654$)
in comparison with the $\Lambda$CDM solutions (\ref{asymLCDM}) (brown dashed lines). }
  \label{F1}
\end{figure}

Regular behaviour in Eq.~(\ref{eqRlim}) at high curvature ${\cal R}$ provides a way for
setting  us possibility the corresponding initial conditions at an arbitrary initial
point $N=N_{ini}$ (or $a_{ini}=e^{ N_{ini}}$)
\begin{equation}
E(N_{ini})=E_{ini}, \qquad {\cal R}(N_{ini})={\cal R}_{ini}\ .
   \label{initial}\end{equation}
Then, by assuming a particular starting point, the system of equations (\ref{eqE}),
(\ref{eqR}) can be integrated and the corresponding free parameters compared to data. As
we start integrating far enough from the present time and close to the CMB
($a_{ini}$ is less than $10^{-3}$, corresponding to the CMB observations), the epoch
 is
the radiation dominated epoch, such that the solutions are assumed to
behave as:
\begin{equation}
{\cal R}\simeq A a^{-4}=A e^{-4N},\qquad E^2\simeq B a^{-4}\ ,
   \label{RElim}\end{equation}
where $A$ and $B$ are two positive constants to be determined by the equations. Then, by
substituting these expressions and their derivatives into Eqs.~(\ref{eqE}) and
(\ref{eqRlim}) and assuming $N=N_{ini}$, the following identities are obtained:
  $$
2\alpha(\Omega_m^* X_r+\alpha\Omega_\Lambda^*A)=\Omega_\Lambda^*A\Psi^2,\qquad
\Psi=1-\alpha\Big(4N_{ini}+3-\log\frac A2\Big),\qquad B=(\Omega_m^* X_r+\alpha\Omega_\Lambda^*A)/\Psi,
 $$
which provides the asymptotical amplitudes $A$ and $B$, and consequently the initial
conditions $E_{ini}$, ${\cal R}_{ini}$ for the system (\ref{eqE}), (\ref{eqR}). An
example is depicted in Fig.~\ref{F1}, where the evolution for $E(a)$ and ${\cal R}(a)$
is shown for the $F(R)$ model (\ref{FRlate}) in comparison with the $\Lambda$CDM
model for $\Omega_m^0=0.2827$ (brown dashed lines). Here we have used the following
 values of parameters from Table.~\ref{Estim}: $\alpha=0.07$, $\beta=1.39$,
$\Omega_m^*=0.2807$, $\Omega_\Lambda^*=0.587$ (dash-dotted blue lines); $\alpha=0.0051$,
$\beta=1.95$, $\Omega_m^*=0.2827$, $\Omega_\Lambda^*=0.654$ (solid red lines); the last
values are optimal if we add restrictions from the CMB data. As shown, the model (\ref{FRlate}) mimics quite well $\Lambda$CDM model for the period
$10^{-5}<a<0.54$.

Note also that the model parameters  (\ref{H0OmOL}) $H_0^*$ and $\Omega_m^*$ do not
coincide in general with the real values of the $F(R)$ model $H_0=H(t_0)$,
$\Omega_m^0=\frac{\kappa^2\rho_m(t_0)}{3(H_0)^2}$, as
 $$
 H_0\ne H^{*}_0, \qquad \Omega_m^0\ne \Omega_m^{*}\ ,
 $$
hold in general for any $F(R)$ models \cite{OdintsovSGS:2017}, since an $F(R)$ model may
recover $\Lambda$CDM at large redshifts, but its corresponding late-time evolution
deviates from $\Lambda$CDM, such that the above quantities as measured today $t=t_0$
would differ from the  $\Lambda$CDM values. Nevertheless, both set of parameters are
connected via the relation of the physical matter density
 \begin{equation}
 \Omega_m^0H_0^2=\Omega_m^{*}(H^{*}_0)^2=\frac{\kappa^2}3\rho_m(t_0),
  \label{H0Omm}\end{equation}
As will be shown below, this remark is important when performing the fitting analysis
for the observable parameters in Sect.~\ref{Data}. In addition, the sum of the
parameters $\Omega_m^*$ and  $\Omega_\Lambda^*$ may not be equal to 1 for the $F(R)$
model (\ref{FRlate}), as the $\Omega_\Lambda^*$ enters in the equations in a completely
different way, unlike for flat $\Lambda$CDM model, where $\Omega_m^0+ \Omega_\Lambda=1$
is satisfied. This fact was discussed and analyzed in Ref.~\cite{OdintsovSGS:2017}. In
the next section, we use the above procedure for integrating the system of equations
(\ref{eqE}), (\ref{eqR}) and apply to the fits with the data.

\section{Data analysis}
\label{Data}

Let us now test the $F(R)$ model (\ref{explog}) and compare its observational
manifestations to recent data from Type Ia supernovae (SNe Ia)
\cite{Union21SNe,Pantheon17}, baryon acoustic oscillations (BAO) \cite{Eisen05},
estimations of the Hubble parameter $H(z)$ \cite{Hdata} and parameters from the cosmic
microwave background radiation (CMB) \cite{Planck13}. All these observations are
connected to redshifts $z\le1100$, such that we work with the expression (\ref{FRlate}),
which describes well the model (\ref{explog}) after inflation. We also fix the
radiation-matter ratio $X_r=\rho_r^0/\rho_m^0$ given in (\ref{Xrm})  (see
\cite{Planck13}), so there are  5 free parameters for our model (\ref{FRlate}):
 \be
 \alpha, \quad\beta, \quad \Omega_m^*, \quad \Omega_\Lambda^* ,\quad H_0^*.
\label{FreeParam} \ee

Recall that in the limit $\alpha=0$ (without  logarithmic corrections) this model transforms into the standard exponential $F(R)$ case \cite{Linder2009} with 4  free parameters, and the $\Lambda$CDM scenario is recovered for $\alpha=0$, $\beta\to+\infty$. In order to fit the model to the observations, here we use the technique of the maximum likelihood.

\subsection{Supernovae Ia data}

Here we use the largest recent SNe Ia catalogue Pantheon sample
\cite{Pantheon17}, which includes $n_{SN}=1048$ data points with redshifts $z_i\in [0 ,\
2.26]$ and distance moduli $\mu_i^{obs}$ of SNe Ia. We also compare the Pantheon data
\cite{Pantheon17} with the Union 2.1 SNe Ia catalogue \cite{Union21SNe} ($n_{SN}=580$
data points).

 For any set of SNe Ia data we estimate differences between  $\mu_i^{obs}$ and
  the corresponding theoretical
values $\mu^{th}(z_i) $, which are logarithms of the luminosity distance  $D_L (z_i)$:
\begin{equation}
\mu (z)\equiv\mu^{th}(z) = 5 \log_{10} \frac{D_L(z)}{10\mbox{pc}}, \qquad D_L (z)= c
(1+z) \int_0^z \frac{d\tilde z}{H (\tilde z)}. \label{mu}
\end{equation}

For our model with different values of the free parameters (\ref{FreeParam}) we
calculate $E(z)$, $H(z)=H_0^*E(z)$, the functions (\ref{mu}) and the corresponding
$\chi^2$ function, which yields
 \begin{equation}
\chi^2_{SN}( \alpha,\beta,\Omega_m^*,\Omega_\Lambda^*)=\min\limits_{H_0^*}
\sum_{i,j=1}^{n_{SN}} \Delta\mu_i\big(C_{SN}^{-1}\big)_{ij} \Delta\mu_j,\qquad
\Delta\mu_i=\mu^{th}(z_i,\alpha,\dots)-\mu^{obs}_i,
  \label{chiSN}\end{equation}
 where  $C_{SN}$ is the $n_{SN}\times n_{SN}$ covariance matrix.
Here we marginalize over $H_0^*$, which is usually considered as a nuisance parameter
for SNe Ia data \cite{OdintsovSGS:2017,Sharov16,PanSh17,SharovBPNC17}. A similar
marginalization is performed for other sources of data.

\subsection{BAO data}


Observational data, connected to baryon acoustic oscillations (BAO), include measurements of two magnitudes \cite{Eisen05}:
 \begin{equation}
 d_z(z)= \frac{r_s(z_d)}{D_V(z)},\qquad
  A(z) = \frac{H_0\sqrt{\Omega_m^0}}{cz}D_V(z),
  \label{dzAz} \end{equation}
where the distances  $ D_V(z)=\Big[cz D_M^2(z)\big/H(z)\Big]^{1/3}$ and
$D_M(z)=D_L(z)/(1+z)$ are expressed via $D_L(z)$  (\ref{mu}).
 The values (\ref{dzAz}) were estimated for definite redshift
ranges of galaxy clusters with mean redshifts $z=z_i$  from a peak in the correlation
function of the galaxy distribution at the comoving sound horizon scale $r_s(z_d)$,
where  $z_d$ corresponds to the end of the baryon drag era.

In this paper we consider 17 BAO data points for $d_z(z)$ and 7 data points for $A(z)$
from Refs.~\cite{BAOdata}, represented here in Table \ref{TBAO}.
 In our calculations with the Union 2.1 SNe Ia catalogue \cite{Union21SNe} (green
 contours in Figs.~\ref{F2}, \ref{F3}) we also included 9 recent BAO $d_z$ data points
 from Ref.~\cite{Wang17}.

\begin{table}[th]
\centering
 {\begin{tabular}{||l|l|l|l|l|l||}
\hline
 $z$  & $d_z(z)$ &$\sigma_d$    & $ A(z)$ & $\sigma_A$   & Survey\\ \hline
 0.106& 0.336  & 0.015 & 0.526& 0.028&  6dFGS \\ \hline
 0.15 & 0.2232 & 0.0084& -    & -    &  SDSS DR7  \\ \hline
 0.20 & 0.1905 & 0.0061& 0.488& 0.016&  SDSS DR7 \\ \hline
 0.275& 0.1390 & 0.0037& -    & -    &  SDSS DR7 \\ \hline
 0.278& 0.1394 & 0.0049& -    & -    & SDSS DR7 \\ \hline
 0.314& 0.1239 & 0.0033& -    & -    &  SDSS LRG \\ \hline
 0.32 & 0.1181 & 0.0026& -    & -    & BOSS DR11 \\ \hline
 0.35 & 0.1097 & 0.0036& 0.484& 0.016& SDSS DR7 \\ \hline
 0.35 & 0.1126 & 0.0022& -    & -    & SDSS DR7 \\ \hline
 0.35 & 0.1161 & 0.0146& -    & -    & SDSS DR7 \\ \hline
 0.44 & 0.0916 & 0.0071& 0.474& 0.034&  WiggleZ \\ \hline
 0.57 & 0.0739 & 0.0043& 0.436& 0.017&  SDSS DR9 \\ \hline
 0.57 & 0.0726 & 0.0014& -    & -    &  SDSS DR11 \\ \hline
 0.60 & 0.0726 & 0.0034& 0.442& 0.020&  WiggleZ \\ \hline
 0.73 & 0.0592 & 0.0032& 0.424& 0.021& WiggleZ \\ \hline
 2.34 & 0.0320 & 0.0021& -& - &  BOSS DR11 \\ \hline
 2.36 & 0.0329 & 0.0017& -& - &  BOSS DR11 \\  \hline
 \end{tabular}
 \caption{BAO data $d_z(z)=r_s(z_d)/D_V(z)$ and $A(z)$ (\ref{dzAz}).}
 \label{TBAO}} \end{table}

Both observed values  (\ref{dzAz})  are independent of the Hubble constant $H_0$: the distances $D_V(z)$ and  $r_s(z_d)$ are proportional to $H_0^{-1}$. For  the sound horizon scale $r_s(z_d)$ we ensure this dependence by using the fitting formula \cite{OdintsovSGS:2017,Sharov16}
 \begin{equation}
 r_s(z_d)=\frac{(r_d\cdot h)_{fid}}h,\qquad
 h=\frac{H_0}{100\mbox{ km}/(\mbox{s}\cdot\mbox{Mpc})}\ .
 \label{rsh}\end{equation}
The best fit  $(r_d\cdot h)_{fid}=104.57\pm1.44$ Mpc was obtained in Ref.~\cite{Sharov16} for the $\Lambda$CDM model. For the product $H_0\sqrt{\Omega_m^0}$
in the expression $A(z)$ we can use the equivalence (\ref{H0Omm}) that may be rewritten
as $H_0\sqrt{\Omega_m^0}=H_0^{*}\sqrt{\Omega_m^{*}}$.

In this approach the $\chi^2$ function for the BAO values (\ref{dzAz}) is
 \begin{equation}
 \chi^2_{BAO}(\alpha,\beta,\Omega_m^*,\Omega_\Lambda^*)=\Delta d\cdot C_d^{-1}(\Delta d)^T+
\Delta { A}\cdot C_A^{-1}(\Delta { A})^T,
  \label{chiB} \end{equation}
 where $\Delta d$ and $\Delta A$ are line elements $\Delta d_i=d_z^{obs}(z_i)-d_z^{th}(z_i,\dots)$ and $\Delta A_i=A^{obs}(z_i)-A^{th}(z_i,\dots)$. The covariance matrices
  $C_{d}$ and $C_{A}$ for correlated BAO data \cite{BAOdata} are described in detail in Ref.~\cite{Sharov16}.

\subsection{$H(z)$ data}

We also include in our analysis estimations of the Hubble parameter $H(z)$  measured by the method of cosmic chronometers, i.e., differential ages $\Delta t$ of galaxies at certain redshifts $z$
\cite{Hdata}. This method uses the relation
 $$ 
 H (z)= \frac{\dot{a}}{a}= -\frac{1}{1+z}\frac{dz}{dt}  \simeq -\frac{1}{1+z}
\frac{\Delta z}{\Delta t}.
 $$ 
Here we use $n_H=31$ values for $H(z)$ estimated with the mentioned method, including 30 data points from Refs.~\cite{Hdata} and Ref.~\cite{Ratsimbazafy17}. For these  data points we calculate the corresponding $\chi^2$ function
\begin{equation}
\chi^2_{H}=\min\limits_{H_0^*} \sum_{i=1}^{n_H} \left[\frac{H^{obs}(z_i)-H^{th}(z_i,
\alpha,\dots)}{\sigma_{H,i}}\right]^2,
\label{chiH} \end{equation}
The Hubble constant is marginalized in the expression for the $\chi^2$, as shown in \cite{OdintsovSGS:2017,SharovBPNC17}. We do not include $H(z)$ estimations from line-of-sight BAO dat \cite{BAOdata} to avoid correlation with the BAO data points taken into account in $\chi^2_{BAO}$ (\ref{chiB}).

\subsection{CMB data}
\label{CMBdata}

Here we use the CMB parameters at the photon-decoupling epoch $z_*=1089.90 \pm0.30$ \cite{Planck13} in the following form
\cite{WangW2013,HuangWW2015}:
 \begin{equation}
  \mathbf{x}=\big(R,\ell_A,\omega_b\big);\qquad R=\sqrt{\Omega_m^0}\frac{H_0D_M(z_*)}c,\quad
 \ell_A=\frac{\pi D_M(z_*)}{r_s(z_*)},\quad\omega_b=\Omega_b^0h^2,
 \label{CMB} \end{equation}
 where the transverse comoving distance $D_M$ and the comoving sound horizon $r_s$ at $z_*$ are
  \begin{equation}
  D_M(z_*)=\frac {D_L(z_*)}{1+z_*} = c \int_0^{z_*}
\frac{d\tilde z}{H (\tilde z)}\ , \quad
 r_s(z)=\frac1{\sqrt{3}}\int_0^{1/(1+z)}\frac{da}
 {a^2H(a)\sqrt{1+\big[3\Omega_b^0/(4\Omega_r^0)\big]a}}\ .
  \label{rs2}\end{equation}
The corresponding distances are given by \cite{HuangWW2015}
  \begin{equation}
  R^{Pl}=1.7448\pm0.0054,\quad \ell_A^{Pl}=301.46\pm0.094,\quad\omega_b^{Pl}=0.0224\pm0.00017,
   \label{CMBpriors} \end{equation}
   with  the covariance matrix
$$C_{CMB}=\|\tilde
C_{ij}\sigma_i\sigma_j\|,\qquad
 \tilde C=\left(\begin{array}{ccc} 1 & 0.53 & -0.73\\ 0.53 & 1 & -0.42\\ -0.73 & -0.42 & 1
\end{array} \right)$$
 from Planck collaboration data \cite{Planck13} with free amplitude of the lensing power spectrum. Here $\Omega_b^0$ is the current baryon density and the sound horizon
$r_s(z_*)$ is calculated by using  Eq.~(\ref{rs2}) and the correction $\Delta r_s=\frac{dr_s}{dz} \Delta z$. The $\chi^2$ function for the data (\ref{CMB}-\ref{CMBpriors})
  \begin{equation}
\chi^2_{CMB}=\min_{H_0^*,\omega_b}\Delta\mathbf{x}\cdot
C_{CMB}^{-1}\big(\Delta\mathbf{x}\big)^{T},\qquad \Delta
\mathbf{x}=\mathbf{x}-\mathbf{x}^{Pl}\ .
 \label{chiCMB} \end{equation}
 includes  marginalizing over the nuisance parameters $\omega_b=\Omega_b^0h^2$ and $H_0^*$.
 Note that the minimum over $H_0^*$ is calculated simultaneously for both $H(z)$ (\ref{chiH}) and CMB (\ref{chiCMB}) data. The results for the $F(R)$ model (\ref{FRlate}) are provided in the next section.

\section{Results}
\label{Analysis}

Here we use the above Pantheon SNe Ia,  $H(z)$, BAO and CMB data to constrain the
parameters for the model  (\ref{explog}). The most strict limitations are produced by
the CMB  data (\ref{chiCMB}), so we analyse separately the $\chi^2$ function as follows:
 \be
  \chi^2_{\Sigma3}=\chi^2_{SN}+\chi^2_H+\chi^2_{BAO}\ ,
 \label{chi3} \ee
which relates the  SNe Ia,  $H(z)$ and BAO observations for the redshift range $0<z\le2.36$. Finally, we compare $\chi^2_{\Sigma3}$ to the total $\chi^2$ including CMB data:
 \be
  \chi^2_{tot}=\chi^2_{SN}+\chi^2_H+\chi^2_{BAO}+\chi^2_{CMB},
 \label{chitot} \ee
 where $\chi^2_{CMB}$ is connected to a redshift $z\simeq 1000$. The free parameters of our $F(R)$ model (\ref{FRlate}) are reduced after marginalizing over $H_0^*$ (and over  $\omega_b$ for  $\chi^2_{CMB}$), such that only 4 free parameters remained: $\alpha$, $\beta$, $\Omega_m^*$ and $\Omega_\Lambda^*$. For the  $\chi^2$ functions (\ref{chi3}) and (\ref{chitot}) we obtain
two-parameter and one-parameter distributions by marginalising over the other parameters.

\begin{figure}[bh]
   \centerline{ \includegraphics[scale=0.66,trim=5mm 0mm 2mm -1mm]{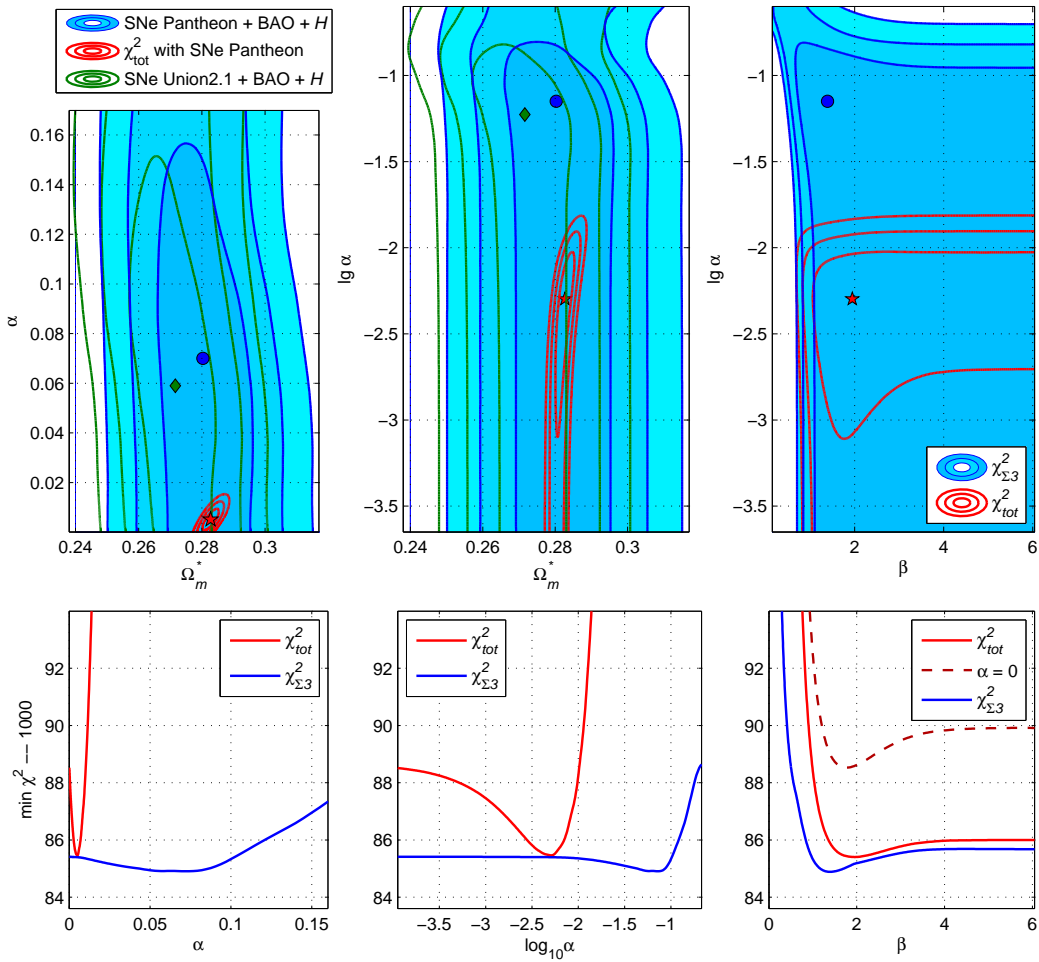}}
\caption{Top panels correspond  to the contour plots for the free parameters of the
model. For the Pantheon SNe Ia data blue regions depict $\chi^2_{\Sigma3}$ (SNe+$H$+BAO)
whereas red lines identify $\chi^2_{tot}=\chi^2_{\Sigma3}+\chi^2_{CMB}$. Green contours
correspond to  $\chi^2_{\Sigma3}$ for the Union 2.1 SNe Ia data.  Bottom panels show the
dependence of the minimum $\chi^2$ for $\Sigma3$ (blue lines) and for the total data
(red lines) for the Pantheon SNe Ia sample. In the bottom-right panel  the limit
$\alpha=0$ for $\chi^2_{tot}$ is also included (the dashed line).}
  \label{F2}
\end{figure}

 Thus, the two-parameter distributions  of $\chi^2_{\Sigma3}$ (filled blue contours)
and $\chi^2_{tot}$ (red contours) the Pantheon SNe Ia dataset \cite{Pantheon17}, are
depicted in the top panels of Fig.~\ref{F2} as contour plots of $1\sigma$, $2\sigma$ and
$3\sigma$ confidence regions. The corresponding one-parameter distributions are in the
bottom panels. In the top-left panels we compare these results with $\chi^2_{\Sigma3}$
distribution for the Union 2.1 SNe Ia data \cite{Union21SNe} ($n_{SN}=580$ data points)
with 9 additional BAO data points from Ref.~\cite{Wang17}, shown as green contours.

  Note that
  in the top-left panel for the $\Omega_m^*-\alpha$ plane we show the
distributions
  \begin{equation}
\chi^2_{\Sigma3}(\alpha,\Omega_m^*)=\min\limits_{\beta,\Omega_\Lambda^*}\chi^2_{\Sigma3}(\alpha,\beta,\Omega_m^*,\Omega_\Lambda^*),\qquad
\chi^2_{tot}(\alpha,\Omega_m^*)=\min\limits_{\beta,\Omega_\Lambda^*}\chi^2_{tot}(\alpha,\beta,\Omega_m^*,\Omega_\Lambda^*).
 \label{2param} \end{equation}

The blue dots for $\chi^2_{\Sigma3}$,  the red stars for $\chi^2_{tot}$ and the green
diamonds for the Union 2.1 SNe Ia data denote the best fits of the corresponding
two-dimensional distributions, which are summarised in Tables~\ref{Estim},
\ref{Estim580}. In these tables the $1\sigma$ errors of the model parameters are
calculated via one-parameter distributions $\chi^2(p_j)$ for the corresponding $\chi^2$
functions and likelihoods ${\cal L}(p_j)$. In particular, for $\chi^2_{tot}$ these
functions are
   \begin{equation}
\chi^2_{tot}(p_j)=\min\limits_{\mbox{\scriptsize other
}p_k}\chi^2_{tot}(p_1,\dots),\qquad {\cal L}_{tot}(p_j)= \exp\bigg[-
\frac{\chi^2_{tot}(p_j)-m_{tot}^{abs}}2\bigg]\ ,
 \label{1param} \end{equation}
 where $p_j$ is the model parameter and the minimum is obtained by marginalising over all
the other free parameters, being $m_{tot}^{abs}$ the absolute minimum for
$\chi^2_{tot}$.

In the bottom panels of Fig.~\ref{F2} we  compare the one-parameter distributions of
$\chi^2_{\Sigma3}$ and $\chi^2_{tot}$ with $p_j=\alpha$ and $\beta$ for datasets
including the Pantheon SNe Ia sample \cite{Pantheon17}.

One can see in Fig.~\ref{F2} and in Table~\ref{Estim} that the best fit leads to
$\alpha=0.070_{-0.070}^{+0.048}$ 
 for $\chi^2_{\Sigma3}$, which is one order of magnitude
larger than $\alpha=0.0051_{-0.0030}^{+0.0027}$ 
for $\chi^2_{tot}$.
  The latter is in better
agreement with the limitation (\ref{alpll1}) which recall that has to be satisfied
during the post-inflationary era ($2\le {\cal R}<R_0/(2\Lambda)$). The best fit for
$\alpha$ is small  for $\chi^2_{tot}$, so the corresponding red contours and lines are
shifted to the margins in the left panels of Fig.~\ref{F2}. Due to this reason, we use
the variable $\lg\alpha\equiv\log_{10}\alpha$ instead of $\alpha$ in other panels. In
particular, in the top-center and right panels, the contour plots depict the
two-dimensional distributions $\chi^2_{\Sigma3}(\alpha,\beta)$ and
$\chi^2_{tot}(\alpha,\beta)$ in the $\lg\alpha-\beta$ plane.

In the bottom-right panel of Fig.~\ref{F2}  the plot (\ref{1param}) of
$\chi^2_{tot}(\beta)$ are  compared with the similar plot
for the case $\alpha=0$ of this model (the dashed lines). 
 Naturally, the presence of the
logarithmic correction with an additional parameter $\alpha$ helps to diminish the
absolute minima for the $\chi^2$ functions, such that the logarithmic corrections in
Eq,~(\ref{FRlate}) provides a better fit than in its absence ($\alpha=0$).

 \begin{table}[ht]
 \centering
 {\begin{tabular}{||l|c||c|c|c|c|l||}  \hline
 Model  & data & $\alpha$ & $\beta$& $\Omega_m^*$& $\Omega_\Lambda^*$ &$ \min\chi^2/d.o.f$  \\ \hline\hline
Exp$\,F(R)+\log$& $\chi^2_{\Sigma3}$& $0.070_{-0.070}^{+0.048}$ & $1.39_{-0.53}^{+\infty}$  & 
$0.2807_{-0.010}^{+0.0102}$  & $0.587_{-0.074}^{+0.106}$ & 1084.90 / 1099\rule{0pt}{1.2em}  \\
 \hline
Exp$\,F(R)$   & $\chi^2_{\Sigma3}$& 0 & $1.88_{-0.66}^{+\infty}$ & 
$0.282_{-0.0095}^{+0.010}$  & $0.654_{-0.059}^{+0.052}$  & 1085.41 / 1100  \rule{0pt}{1.2em}  \\
 \hline
$\Lambda$CDM & $\chi^2_{\Sigma3}$& 0  & $\infty$  & 
$0.2859^{+0.0089}_{-0.009}$  & $0.714_{-0.009}^{+0.009}$ & 1087.16 / 1102 \rule{0pt}{1.2em} \\
 \hline \hline
 Exp$\,F(R)+\log$& $\chi^2_{tot}$& $0.0051_{-0.0030}^{+0.0027}$ & $1.95_{-0.70}^{+\infty}$ &
$0.2827_{-0.0018}^{+0.0017}$  & $0.654_{-0.046}^{+0.017}$ & 1085.41 / 1102 \rule{0pt}{1.2em}  \\
 \hline
Exp$\,F(R)$ & $\chi^2_{tot}$& 0  & $1.76_{-0.49}^{+1.33}$&
$0.2803_{-0.001}^{+0.001}$  & $0.655_{-0.042}^{+0.014}$   & 1088.53 / 1103 \rule{0pt}{1.2em} \\
  \hline
$\Lambda$CDM & $\chi^2_{tot}$& 0 & $\infty$ &
$\!0.2807_{-0.0004}^{+0.0003}\!$  & $\!0.7193_{-0.0003}^{+0.0004}\!$   & 1088.91 / 1105 \rule{0pt}{1.2em} \\
 \hline \end{tabular}
\caption{Predictions of the exponential $F(R)$ model with logarithmic corrections (\ref{FRlate}), its analog without
 corrections ($\alpha=0$) and the $\Lambda$CDM model
for the Pantheon SNe Ia data with $H(z)$ and BAO from Table~\ref{TBAO}
($\chi^2_{\Sigma3}=\chi^2_{SN}+\chi^2_H+\chi^2_{BAO}$) and SNe
Ia${}+H(z)+{}$BAO${}+{}$CMB ($\chi^2_{tot}=\chi^2_{\Sigma3}+\chi^2_{CMB}$): $\min\chi^2$
and
  $1\sigma$ estimates of model parameters.}
 \label{Estim}}\end{table}

\begin{figure}[th]
   \centerline{ \includegraphics[scale=0.66,trim=5mm 0mm 2mm -1mm]{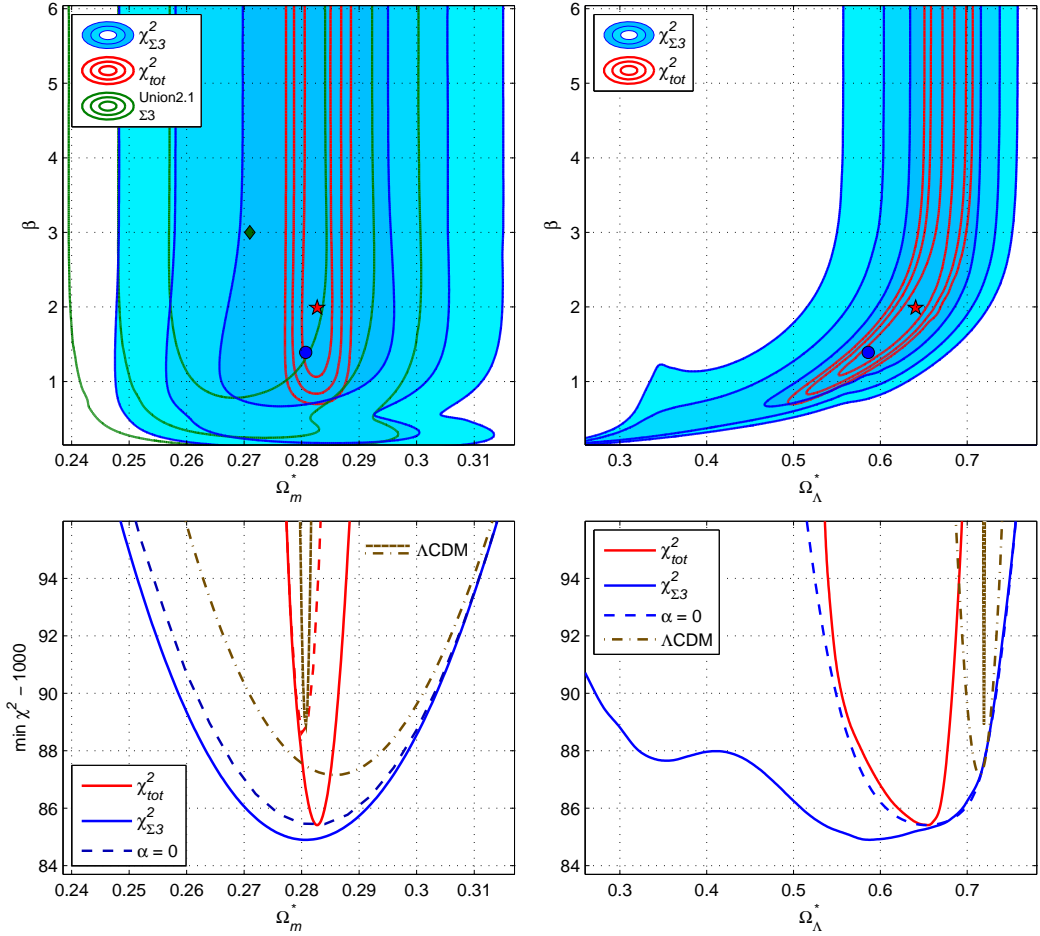}}
\caption{Top panels show the contours  for $\chi^2_{\Sigma3}$ (blue) and  for
$\chi^2_{tot}$ (red) for the Pantheon SNe Ia sample in  the $\Omega_m^*-\beta$  and
$\Omega_\Lambda^*-\beta$ planes.
 In the bottom panels the corresponding one-dimensional distributions are plotted in comparison to
the case without logarithmic corrections $\alpha=0$ (the dashed lines with correspondent
color) and the $\Lambda$CDM model
 (brown dashed lines for $\chi^2_{tot}$ and dash-dotted lines for $\chi^2_{\Sigma3}$).}
  \label{F3}
\end{figure}

These absolute minimum for $\chi^2_{\Sigma3}$ and $\chi^2_{tot}$  are written in the
right column of  Table~\ref{Estim}. Here the degrees of freedom (d.o.f.) are the total
number of data points minus the number of independent model parameters. In Fig.~\ref{F3}
one can see the contour plots for $\chi^2_{\Sigma3}$ and $\chi^2_{tot}$ in the
$\Omega_m^*-\beta$  and  $\Omega_\Lambda^*-\beta$ planes (the top panels). The bottom
panels show how one-dimensional distributions (\ref{1param}) depend on $\Omega_m^*$ and
$\Omega_\Lambda^*$ in comparison to the same model without logarithmic corrections
($\alpha=0$) and the $\Lambda$CDM model. For all these models we observe essentially
more sharp dependence on $\chi^2_{tot}(\Omega_m^*)$ than on
$\chi^2_{\Sigma3}(\Omega_m^*)$. For the flat $\Lambda$CDM model its parameters
$\Omega_m^0$ and $\Omega_\Lambda=1-\Omega_m^0$ are used along the abscissa axes,
 recall that  they differ from $\Omega_m^*$  and  $\Omega_\Lambda^*$
 for the considered $F(R)$ model (\ref{explog}).

One can see in the bottom panels of Fig.~\ref{F3} and in  Table~\ref{Estim} that the
minima of $\chi^2_{\Sigma3}$ and $\chi^2_{tot}$ for the $F(R)$ model are the least in
comparison with the case $\alpha=0$ and the $\Lambda$CDM model.


Similar calculation with the Union 2.1 SNe Ia  dataset and 9 additional BAO data
points from Ref.~\cite{Wang17} are presented in Table~\ref{Estim580} and illustrated in
the top-left panel of Fig.~\ref{F3} (the green contours for $\chi^2_{\Sigma3}$). Note
that the BAO data points from Ref.~\cite{Wang17} bring more essential contribution in
differences of the estimated values in Tables~\ref{Estim} and \ref{Estim580} than the
SNe Ia  data sample  Union 2.1  or Pantheon.

 \begin{table}[ht]
 \centering
 {\begin{tabular}{||l|c||c|c|c|c|l||}  \hline
 Model  & data & $\alpha$ & $\beta$& $\Omega_m^*$& $\Omega_\Lambda^*$ &$ \min\chi^2/d.o.f$  \\ \hline\hline
Exp$\,F(R)+\log$& $\chi^2_{\Sigma3}$& $0.059_{-0.056}^{+0.046}$ & $3.0_{-1.80}^{+\infty}$  & 
$0.271_{-0.0093}^{+0.0089}$  & $0.637_{-0.095}^{+0.043}$ & 588.10 / 640 \rule{0pt}{1.2em}  \\
 \hline
Exp$\,F(R)$   & $\chi^2_{\Sigma3}$& 0 & $5.61_{-4.01}^{+\infty}$ & 
$0.274_{-0.008}^{+0.008}$  & $0.674_{-0.068}^{+0.017}$  & 589.09 / 641 \rule{0pt}{1.2em}  \\
 \hline \hline
 Exp$\,F(R)+\log$& $\chi^2_{tot}$& $0.0014_{-0.0014}^{+0.0025}$ & $4.71_{-2.87}^{+\infty}$ &
$0.2823_{-0.0021}^{+0.0017}$  & $0.661_{-0.049}^{+0.011}$ & 590.27 / 643 \rule{0pt}{1.2em}  \\
 \hline
Exp$\,F(R)$ & $\chi^2_{tot}$& 0  & $3.98_{-2.46}^{+\infty}$&
$0.2814_{-0.0008}^{+0.001}$  & $0.660_{-0.067}^{+0.012}$   & 590.75  / 644 \rule{0pt}{1.2em} \\
 \hline \end{tabular}
\caption{Estimations of model parameters for the $F(R)$ model (\ref{FRlate}) and its
analog without corrections ($\alpha=0$) with the Union 2.1 SNe Ia  data set and BAO data
including Table~\ref{TBAO} and 9 additional data points from Ref.~\cite{Wang17}.}
 \label{Estim580}}\end{table}

\section{Inflationary Era}
  \label{Inflation}

The $\gamma(R) R^2$ term with the logarithmic correction can explain well the early-time inflation, when $R\ge R_0\sim 10^{85}\Lambda$ \cite{OdintsovOS:2017log}. This  inflationary era was investigated in
Ref.~\cite{OdintsovOS:2017log} in the constant-roll inflation description and  a viable inflationary scenario was obtained. Here we pretend to show that model (\ref{explog}) reproduces slow-roll inflation and provides the correct values for the spectral index and the tensor-to-scalar ratio. As pointed our above, during the inflationary epoch $ \beta{\cal R}\gg 1$ holds and the Lagrangian  (\ref{explog}) takes the form:
dominating $\gamma(R) R^2$ term:
 \begin{equation}
 F(R)= R-2\Lambda+\alpha R\log\left(\frac{ R}{4\Lambda}\right)+\gamma_0\left( 1+\gamma_1\log 
 \frac{R}{R_0} 
  \right)R^2.
\label{FRInfl}\end{equation}

An unstable de Sitter point $R= R_{dS}$, corresponding to inflation, arises naturally under the condition \cite{nostriexp}
 \begin{equation}
  G(R_{dS})=0
 \label{Geq0}\end{equation}
where $  G(R)=2F(R)-RF_R$. For the action (\ref{FRInfl}), this condition yields:
 \begin{equation}
\gamma_0\gamma_1 R_{dS}=1-\frac{4\Lambda}{R_{dS}}
+\alpha\left(\log\frac{R_{dS}}{4\Lambda}-1\right).
 \label{RdS}\end{equation}
During the inflationary era we can neglect the term $4\Lambda/R\simeq 10^{-85}$ and solve approximately this equation for  $\alpha\ll 1$, in agreement with the limitation (\ref{alpll1})) and the contraints obtained in the previous section:
 \begin{equation}
  R_{dS}\simeq \frac1{\gamma_0\gamma_1}\Big[1-\alpha\log(4\e\gamma_0\gamma_1\Lambda)\Big].
 \label{RdS1}\end{equation}
As shown, the contribution of the logarithmic term in front of the factor $\alpha$ is not completely negligible. Let us describe the slow-roll inflation for the above  model. The $F(R)$ action can be expressed in terms of a scalar field $\phi$ ~\cite{OdintsovSGS:2017}
 \be
 S=\int {\rm d}^4x\sqrt{-g}\left[\varphi
R-V(\varphi)\right]\ ,
 \label{scalar1} \ee
where the scalar field and its potential are related to the $F(R)$ function through the relations:
 \be
 \phi=F_{R}\ , \qquad V(\phi)=RF_R-F\ .
 \label{phi}
 \ee
The action (\ref{scalar1}) can be transformed into the Einstein frame via the conformal transformation
 $$
\tilde{g}_{\mu\nu}=\phi g_{\mu\nu}\ ,
 $$
 which transforms the action to the Einstein frame, leading to:
  \be \tilde{S}=\int
{\rm
d}^4x\sqrt{-\tilde{g}}\left[\frac{\tilde{R}}{2\kappa^2}-\frac{1}{2}\partial_{\mu}\tilde{\phi}\,\partial^{\mu}\tilde{\phi}-\tilde{V}(\tilde{\phi})\right]\ .
\label{scalar5}
\ee
 Here, we have  redefined the scalar field and the potential as:
 \be \phi=\e^{\sqrt{\frac{2}{3}}\kappa\tilde{\phi}}\ , \qquad
\tilde{V}=\frac{\e^{-2\sqrt{\frac{2}{3}}\kappa\tilde{\phi}}}{2\kappa^2} V\ .
\label{tildephi} \ee
The scalar field mimics an effective cosmological constant during slow-roll inflation,
what is equivalent to the conditions $H\dot{\tilde{\phi}}\gg \ddot{\tilde{\phi}}$ and
$\tilde{V}\gg \dot{\tilde{\phi}}^2$, which can be  expressed in terms of  the 
slow-roll parameters
 \be
  \epsilon= \frac{1}{2\kappa^2} \left(
\frac{\tilde{V}'(\tilde{\phi})}{\tilde{V}(\tilde{\phi})} \right)^2\, ,\qquad \eta=
\frac{1}{\kappa^2} \frac{\tilde{V}''(\tilde{\phi})}{\tilde{V}(\tilde{\phi})}\,.
\label{epseta} \ee While the number of e-folds can be expressed as: \be N  \equiv
\int_{t_{start}}^{t_{end}} {H} {\rm d}t \simeq -\kappa^2
\int_{\tilde{\phi}_{start}}^{\tilde{\phi}_{end}}
\frac{\tilde{V}(\tilde{\phi})}{\tilde{V}'(\tilde{\phi})} {\rm d}\phi\ .\label{Nefolds}
\ee By the relations (\ref{tildephi}), the slow-roll parameters can be expressed in
terms of the the Ricci scalar $R$ for the Lagrangian (\ref{FRInfl}):
 \bea
  \epsilon&=& \frac13  \left(\frac{2F(R)-RF_R}{RF_R-F}\right)^2\simeq
 \frac13 \left[\frac{1+\alpha \log\frac{ R}{4\e\Lambda}+\gamma_0\gamma_1R}
{\alpha +\gamma_0R\,\big(1+\gamma_1+\gamma_1\log\frac{R}{R_0}\big)}\right]^2\,,
 \label{epsilon} \\
 \eta&=&2\epsilon + \frac23 \frac{F_R}{F_{RR}}\frac{d}{dR}\frac{F}{RF_R-F}\,.
\label{eta} \eea
 Here we have assumed the limit $\Lambda/R\simeq 10^{-85}$. During the inflationary period the slow-roll parameters (\ref{epseta}) should satisfy the limitations $\epsilon\ll 1$ and $\eta<1$, while $\epsilon\gtrsim 1$ at the end of
inflation. The slow-roll  parameters $\epsilon$ and $\eta$ are related to the spectral index $n_\mathrm{s}$ of the scalar perturbations originated during inflation and the tensor-to-scalar ratio $r$ as follows:
  \be
n_\mathrm{s} = 1 - 6 \epsilon + 2 \eta\, ,\qquad r = 16 \epsilon \,.
 \label{nsr} \ee
 The last data from Planck and Bicep2 collaborations \cite{Planck-Inflation} constrains the values of the spectral index and the tensor-to-scalar ratio as follows,
\begin{equation}
n_\mathrm{s}=0.968\pm0.006\ ,\qquad r<0.07\ .
  \label{Planck}
\end{equation}

From the relations (\ref{phi}), the scalar field and its potential can be written in terms of the Ricci scalar as follows:
\bea
\phi &=&1+\alpha+\gamma_0\left[2+\gamma_1\left(1+2\log\frac{R}{R_0}\right)\right]R+\alpha\log\left(\frac{R}{4\Lambda}\right)\ , \nn
V(\phi)&=&\left[\alpha+\gamma_0\left[1+\gamma_1\left(1+\log\frac{R}{R_0}\right)\right]R\right]R \ .
\label{phiAndPotential}
\eea
These relations are not analytically invertible, such that we can not obtain an analytical form for the scalar potential (\ref{tildephi}) and consequently for the spectral index and the tensor-to-scalar ratio in terms of the number of e-folds (\ref{Nefolds}), but numerical resources are required, as shown below. Nevertheless, a first qualitative analysis of the model (\ref{FRInfl}) can be carried out by assuming $R\sim R_0$ at the end of inflation and $\alpha\ll 1$, such that the relations (\ref{phiAndPotential}) can be approximated as follows:
\bea
\phi &\sim&1+\gamma_0\left[2+\gamma_1\left(1+2\log\frac{R}{R_0}\right)\right]R\ , \nn
V(\phi)&\sim& \gamma_0\left[1+\gamma_1\left(1+\log\frac{R}{R_0}\right)\right] R^2\ .
\label{AproxphiAndPotential}
\eea
Then, the potential in the Einstein frame (\ref{tildephi}) yields approximately:
 \be
\tilde{V}(\tilde{\phi})\sim
\frac{1}{2\kappa^2}\frac{1+\gamma_1}{\gamma_0(2+\gamma_1)^2}\left(1-\e^{-\sqrt{\frac{2}{3}}\kappa\tilde{\phi}}\right)^2\
. \label{PotentialFinal}
 \ee
This is the potential for the $R^2$ Starobinsky model, such that the appropriate predictions can be achieved. \\

However, here we make a full numerical analysis to obtain reliable information about the
viability of our model. In order to compare the constraints and predictions of our
model, we fix the parameter $R_0= 10^{85}\Lambda$ as estimated in
Ref.~\cite{OdintsovOS:2017log}. Note that the predictions weakly depend on a choice of
$R_0$ because of logarithms (and the small factor $\alpha$ in some cases). Thus, we can
assume two free parameters for the model during the inflationary era: $\gamma_0$ and
$\gamma_1$. It is convenient to introduce (in addition to $\gamma_1$) the dimensionless
parameter
 $$
 \Gamma_0=\gamma_0R_0.
 $$

\begin{figure}[th]
   \centerline{ \includegraphics[scale=0.7,trim=5mm 0mm 2mm -1mm]{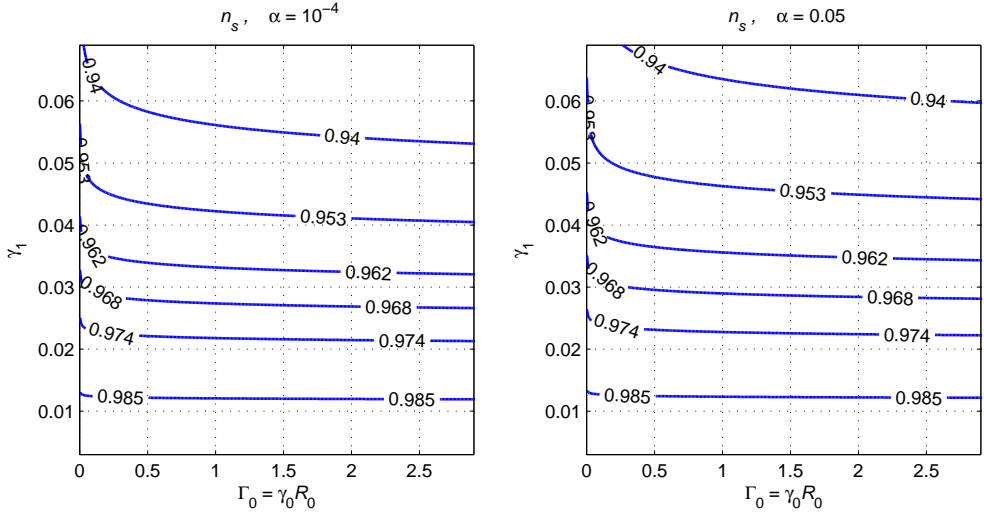}}
\caption{The scalar spectral index $n_\mathrm{s}$ in the  $\Gamma_0-\gamma_1$ plane for
$\alpha=10^{-4}$ (left) and $\alpha=0.05$ (right panel).}
  \label{F4}
\end{figure}

In our numerical calculations for fixed values of the mentioned parameters $\gamma_1$
and $ \Gamma_0$  (and also the third parameter $\alpha$) 
we determine the de Sitter value $R= R_{dS}$ or $R_{dS}/R_0$ from the equation
(\ref{RdS1}) or (\ref{RdS}). For $R= R_{dS}$ we calculate the slow-roll parameters
(\ref{epsilon}), (\ref{eta}),  the spectral index and the scalar-to-tensor ratio
(\ref{nsr}). The calculated  distributions of the spectral index $n_\mathrm{s}$ in the
$\Gamma_0-\gamma_1$ plane are shown in Fig.~\ref{F4} as contour plots (level lines) for
fixed values  $\alpha=10^{-4}$ (left) and $\alpha=0.05$ (the right panel).

One can see that  $n_\mathrm{s}$ weakly depends on the parameters $\Gamma_0=\gamma_0R_0$
and $\alpha$ (for allowed small $\alpha$ values), but essentially depends on $\gamma_1$.
However, the model satisfies the Planck restrictions (\ref{Planck}) in the range
$0.022<\gamma_1<0.032$,  $\Gamma_0>0.5$ and small  $\alpha$. The calculated ratio $r$
(\ref{nsr}) satisfies the limitations (\ref{Planck}) for reasonable values of
parameters. Here we have also required an adequate number of e-foldings $N \simeq 55 -
65$ during the inflationary era.

 \section{Newton's law corrections in $F(R)$ gravity}
\label{Newton}

Extensiones of General Relativity may induce large corrections on the Newton's law at local scales, as at the Earth or the Solar System. $F(R)$ gravities carry an extra scalar degree of freedom that may violate local gravity tests unless the scalar mode is conveniently screened through the so-called chameleon mechanism (see \cite{Khoury:2003rn}), applied to $F(R)$ gravities \cite{Hu:2007nk,Nojiri:2007cq}. In order to show this point, let us start by writing the trace of the field equations (\ref{fieldeq}):
\be
\Box f_R-\frac{1}{3}\left(R+2 f-f_R R\right)=\frac{\kappa^2}{3}T\ ,
\label{trace}
\ee
where we have defined $f(R)=F(R)+R$. Such equation can be interpreted as the equation of motion for the scalar field inherent to $F(R)$ gravities, which is identified by the first derivative $f_R$. In addition, we can define the effective potential for the scalar field as follows:
\be
\frac{\partial V_{eff}}{\partial f_R}=\frac{1}{3}\left(R+2 f-f_R R+\kappa^2T\right)\ .
\label{potential}
\ee
Hence, the effective mass for the scalar field, which plays a fundamental role on the growth of perturbations around a particular solution, is given by the second derivative of the effective potential (\ref{potential}):
\be
m^2_{eff}= \frac{\partial^2 V_{eff}}{\partial f_R^2}=\frac{1}{3}\left(\frac{1+f_R}{f_{RR}}-R\right)\ .
\label{mass}
\ee
In order to avoid tachyons, the effective mass (\ref{mass}) should be positive everywhere, while should be large enough at local scales in order to avoid corrections on the Newton's law \cite{Joyce:2014kja,Hu:2007nk,Nojiri:2007cq}. Let us now investigate the case of the action (\ref{explog}). Since the scales where large corrections on the Newton's law may be induced can be taken as $2\Lambda<<R_{\text{local}}<<R_0$, where recall that $R_0$ is the curvature at the end of inflation, the action (\ref{explog}) can be approximated as follows:
\be
f(R)\sim -2\Lambda \left(1-\alpha \frac{R}{2 \Lambda}\log\frac{R}{4\Lambda}\right)\ .
\label{approxAction}
\ee
Then, the mass of the scalaron leads to:
\be
m^2_{eff}\sim R\left(\frac{1}{\alpha}+\log\frac{R}{4\Lambda}\right)\ .
\label{massapprox}
\ee
Then, the mass is always positive $m^2_{eff}>0$, and for sufficient large curvature $R>>4\Lambda$, the mass is large enough to avoid corrections on the Newton's law at local scales, where the free parameters of the model do not play any role as far as $\alpha>0$, which is satisfied as pointed by the fits shown in Table \ref{Estim}. Also $R>4\Lambda$, but this is obvious at least at scales where the curvature is much larger than the curvature of the universe. \\

However, the so-called matter instability (see \cite{matterInstability}) may also be present at systems where the curvature is large enough, as in the Earth. In order to avoid such instability, we can analyse the equation for the scalar field (\ref{trace}), which can be expressed as follows:
\be
\Box R+\frac{f_{RRR}}{f_{RR}}\nabla_{\mu}R\nabla^{\mu}R+R\frac{1+f_R}{3f_{RR}}-2\frac{R+f}{3f_{RR}}=\frac{\kappa^2}{6 f_{RR}}T\ .
\label{scaleq2}
\ee
We can consider the solution $R=R_e=-\frac{\kappa^2}{2}T$, and a perturbation around such solution $\delta R$. Then, the equation for the perturbation leads to:
\be
\left(\frac{\partial}{\partial t}-U(R_e)\right)\delta R=0\ ,
\label{matterins}
\ee
where
\be
U(R_e)=\left(\frac{F_{RRRR}}{F_{RR}}-\frac{F_{RRR}^2}{F_{RR}^2}\right)\nabla_{\mu}R\nabla^{\mu}R+\frac{R}{3}-\frac{F_RF_{RRRR}R}{3F_{RR}}-\frac{f_{R}}{3F_{RR}}+\frac{2FF_{RRRR}}{3F_RR^2}-\frac{F_{RRRR}R}{3F_{RR}}\ .
\label{pontMattIns}
\ee
In order to avoid exponential growth of the perturbation, the potential (\ref{pontMattIns}) should be negative, $U(R_e)<0$. By introducing the action (\ref{approxAction}) in (\ref{pontMattIns}), the following expression for the potential is obtained:
\be
U(R_e)=\frac{R_e(-1+\alpha)+4\Lambda-2\alpha R_e\log\frac{R_e}{4\Lambda}}{3\alpha}\ .
\label{pont1}
\ee
Hence, the potential is negative in general and particularly for large curvature regimes $R_e>>2\Lambda$, such that the possible perturbations around the solution $R=R_e$ can turn out negligible, and the whole action (\ref{explog}) is suitable also for describing local gravity systems. 

\section{Constraints from Big Bang nucleosynthesis} \label{BBN}

The considered $F(R)$ model (\ref{explog}) shows a significant difference from
the $\Lambda$CDM model along the period before the recombination epoch $a<10^{-3}$. Particularly, during the Big Bang nucleosynthesis (BBN) period at $10^{-9}\le
a\le10^{-8}$. Hence, for our model we should take into account possible restrictions
coming from BBN. One can see in Fig.~\ref{F1} that essentially model (\ref{explog}) differs on the behaviour of the Ricci scalar for
$a<10^{-5}$, whereas the normalized Hubble parameter $E(a)=H/H_0$  behaves  like $E\sim
a^{-2}$ (radiation dominated era) for both the $\Lambda$CDM model as the $F(R)$ model.
The relation $E/E_{\Lambda CDM}$ is close to a constant for $a<10^{-5}$. As shown below,  $E/E_{\Lambda CDM}$ depends mainly on $\alpha$ and tends to 1 when
$\alpha$ goes to zero.\\

During the BBN period $10\le t\le 10^{3}$ sec, the baryon to photon ratio $\eta$, the effective number of neutrinos $N_{eff}$, etc. 
have a direct influence on the resulting abundances of deuterium \cite{CopiDK2004}, helium ${}^4$He
\cite{Planck13,Steigman} and other light elements \cite{PArthENoPE2018}. For our
estimations, we assume the approximated formula for the helium ${}^4$He mass fraction
$Y_p=4n_{\mbox{\scriptsize He}}\big/(n_p+n_n)$ (see Refs.~\cite{Steigman}):
 \be
Y_p=\frac{4n_{\mbox{\scriptsize He}}}{n_p+n_n}=0.2485\pm\Delta Y_p +0.0016(\eta_{10}- 6)
+0.16\bigg(\frac H{H_{\Lambda CDM}}\bigg|_{BBN}-1\bigg)\ .
 \label{Yp} \ee
Here $\eta_{10}=10^{10}\eta$, $A\big|_{BBN}$ means $A(a_{BBN})$; the error in
Refs.~\cite{Steigman} is $\Delta Y_p=0.0006$, however the latest Planck estimation
\cite{Planck13} $Y_p=0.249_{-0.026}^{+0.025}$ and $Y_p=0.2449\pm0.0040$ from
Ref.~\cite{PArthENoPE2018} yield larger values of $\Delta Y_p$.

By considering $\eta_{10}$ as a free parameter in every cosmological model (with the recent
estimation $\eta_{10}=6.13\pm0.13$ \cite{PArthENoPE2018}), we can evaluate the following
limitation for the last term in Eq.~(\ref{Yp}):
 \be
\bigg|\Big(\frac{H}{H_{\Lambda CDM}}\Big)\Big|_{BBN}-1\bigg|=
\bigg|\frac{H_0^*}{H_0}\Big(\frac{E}{E_{\Lambda CDM}}\Big)\Big|_{BBN}-1\bigg|\le
\frac{\Delta Y_p}{0.16}\ .
 \label{limYp} \ee

During the BBN stage the evolution of $E(a)$ and ${\cal R}(a)$ does not depend on the
parameter $\beta$ (because of the factor $e^{-\beta{\cal R}}$ is negligible) and we can
express the ratio $E/E_{\Lambda CDM}$ from the relation (\ref{eqRlim}) as follows:
\be
\frac{E^2_{\Lambda CDM}}{E^2}\bigg|_{BBN}\simeq 1+\alpha\bigg(\frac{d\log{\cal R}}{dN}
 -\Omega_\Lambda^*\frac{\cal R}{E^2}+1+\log\frac{\cal R}2\bigg)\bigg|_{BBN}=
 1+\alpha\big(Q+\log{\cal R}|_{BBN}\big)\;.
   \label{ELE}\ee
This expression describes the mentioned behaviour of  $E/E_{\Lambda CDM}$ in the left panel of Fig.~\ref{F1}.  The value
 $$
Q=\frac{d\log{\cal R}}{dN}-\Omega_\Lambda^*\frac{\cal R}{E^2}+1-\log2=-3.661\pm0.008
 $$
 weakly depends on the model parameters $\Omega_m^*$, $\Omega_\Lambda^*$, because
$d\log{\cal R}/dN\simeq-4$ from Eq.~(\ref{RElim}) and the ratio ${\cal R}/{E^2}$ is
small (close to $10^{-2}$) for the parameters from Table.~\ref{Estim}. For the logarithm
 $$
\log{\cal R}|_{BBN}=64.94\pm4.61
 $$
 the errors are larger, they are determined by the duration of the BBN period (1 order of magnitude for $a$, corresponding to
4 for $\log_{10}R$).
 Thus, we can estimate the expression
\begin{equation}
Q+\log{\cal R}|_{BBN}=61.28\pm4.62.
   \label{QlnR}\end{equation}

Hence, the parameter $\alpha$ should be small enough to satisfy the limitation (\ref{alpll1})
$\alpha\big[1+\log({\cal R}/2)\big]\ll 1$, which for the BBN period may be rewritten as
 \be
 \alpha\big(Q+\log{\cal R}|_{BBN}\big)\ll 1\;.
 \label{alpll2}\ee
 Under this condition we can express
 $ 
 \big(E/E_{\Lambda CDM}\big)\big|_{BBN}\simeq
 1-\frac12\alpha \big(Q+\log{\cal R}|_{BBN}\big)
 $
from Eq.~(\ref{ELE}), substituting it in the inequality (\ref{limYp}) and obtain the
following BBN restriction for $\alpha$:
   \begin{equation}
\alpha\le\frac2{Q+\log{\cal R}|_{BBN}}\bigg(\frac{\Delta
Y_p}{0.16}\frac{H_0}{H_0^*}+\frac{H_0^*-H_0}{H_0^*}\bigg).
   \label{alphaBBN}\end{equation}
Then,  the constraints depend on a choice of the $F(R)$ model parameter, as well as $H_0^*$ (\ref{H0OmOL}) 
may differ from  $H_0^{\Lambda CDM}$. However, for $H_0^*\ge
H_0$ or $H_0^*\simeq H_0$, the constraint (\ref{alphaBBN}) will be fulfilled as a
consequence of the condition (\ref{alpll1}) or  (\ref{alpll2}), if we use the latest
Planck estimation \cite{Planck13} $\Delta Y_p=0.026$. For this $\Delta Y_p$, Eq.~
(\ref{QlnR}), under the simplest assumption $H_0^*=H_0$ is reduced to
$$\alpha\le0.0053\;.$$ This condition is  fulfilled, if we take the best fitted value $\alpha=0.0051$ for
$\chi^2_{tot}$ from Table~\ref{Estim} and also for the corresponding $\alpha=0.0014$ from Table~\ref{Estim580} with SNe Ia data \cite{Union21SNe}.

\section{Conclusions}
\label{conclusions}
%
Along the present manuscript, we have focused on a deep analysis of a particular $F(R)$ model. As pointed out in the vast literature about this class of modified gravities, $F(R)$ gravity can reproduce well an accelerating expansion, leading to a possible solution to the dark energy problem as well as to a consistent description of the inflationary paradigm. However, this type of extensions of GR carry an additional scalar mode that may affect the well known predictions of GR at local scales and lead to possible ghost modes. Nevertheless, here we have focused on a particular type of $F(R)$ gravities, the so-called exponential gravity, that is able to satisfy the basic conditions for its viability \cite{Pogosian:2007sw}. Then, we have considered a logarithmic correction in order to provide a test for this type of $F(R)$ theory and check how a deviation is allowed, also in comparison to the $\Lambda$CDM model. As shown in this manuscript, the aim for considering such a class of logarithmic corrections lies on the fact that the new theory still satisfies the viability conditions (under some conditions of the free parameter) and provides an extra term in the action that evolutes smoothly along the cosmological evolution (far from the pole obviously). Then, by using several datasets covering redshifts from the CMB till $z=0$, we have obtained the corresponding constraints and best fits for the parameters. As shown in Figs.~\ref{F1} and \ref{F2}, CMB data provides a much stronger constraint on the free parameters. In addition, the presence of the logarithmic correction, modelled by the free parameter $\alpha$, leads to a better fit than in absence of the logarithmic correction and also better than the $\Lambda$CDM model, obviously at the price of introducing an additional degree of freedom. Moreover, the parameter $\beta$ leads to the same natural result as when testing exponential gravity without any corrections \cite{OdintsovSGS:2017}, i.e. the absence of an upper bound, as $\Lambda$CDM is recovered for $\beta\rightarrow\infty$. Finally, another remarkable result corresponds to the strong constraints obtained on the energy densities in comparison to $\Lambda$CDM model, what provides a better way to test this type of theories. \\

In addition, $R^2$ inflation has also been studied at the end of the paper, where a logarithmic correction is also considered. Such type of analysis provides a way for testing deviations from $R^2$ inflation, which is considered nowadays as one of the best models that satisfies the constraints on the growth of scalar and tensor perturbations during inflation, such that any correction to $R^2$ model may provide information about how far one can go away, specially for the incoming data in the future \cite{delaCruz-Dombriz:2016bjj}. Hence, we have analysed the inclusion of a logarithmic correction, which essentially recovers the $R^2$ predictions for the appropriate limits. A full numerical analysis is also performed, where the possible deviations, managed by the parameter $\gamma_1$, are allowed but kept small, as shown in Fig.~\ref{F4}. However, the theoretical constraints obtained on the parameters are in agreement to the constraints from Planck data, keeping the inflationary part of the model as a reliable one.\\

Finally, as shown in section \ref{Newton}, this particular $F(R)$ model
avoids the presence of large corrections on the Newton's law as well as the appearance
of large instabilities at local systems, leading to a suitable model that recovers the
well known results of GR at the appropriate scales. The model also satisfies the BBN
constraint (\ref{alphaBBN}) on its parameter $\alpha$, 
if $\alpha$ obeys the restriction (\ref{alpll1}). In particular, this is true for the
best fitted value $\alpha=0.0051$ from Table~\ref{Estim} for the total set of
observational data (SNe Ia, $H(z)$, BAO and CMB). \\

 Hence, after this deep analysis, logarithmic corrections are established as potential viable terms in this $F(R)$ model, reproducing both the dark energy epoch as the inflationary phase.

\section*{Acknowledgments}
%
SDO and DSG acknowledge the support by MINECO (Spain), project FIS2016-76363-P.
DSG is also funded by the grant No.~IT956-16 (Basque Government, Spain). This article is based upon work from CANTATA COST (European Cooperation in Science and Technology) action CA15117,  EU Framework Programme Horizon 2020.


\begin{thebibliography}{}
%

\bibitem{reviews1}
  Y.~F.~Cai, S.~Capozziello, M.~De Laurentis and E.~N.~Saridakis,
  Rept.\ Prog.\ Phys.\  {\bf 79}, no. 10, 106901 (2016)
  doi:10.1088/0034-4885/79/10/106901
  [arXiv:1511.07586 [gr-qc]];

  S.~Nojiri, S.~D.~Odintsov and V.~K.~Oikonomou,
   Phys.\ Rept.\  692 (2017)  1, 
  arXiv:1705.11098. 

S. Nojiri and  S.D. Odintsov,
   Phys.\ Rept.\  {\bf 505}, 59 (2011);

S. Nojiri and S.D. Odintsov,
  eConf {\bf C0602061}, 06 (2006)
  [Int.\ J.\ Geom.\ Meth.\ Mod.\ Phys.\  {\bf 4}, 115 (2007)].

 S. Capozziello and M. De Laurentis,
   Phys.\ Rept.\  {\bf 509}, 167 (2011);

  A.~de la Cruz-Dombriz and D.~Saez-Gomez,
  Entropy {\bf 14}, 1717 (2012)
  doi:10.3390/e14091717
  [arXiv:1207.2663 [gr-qc]].


\bibitem{Capozziello2002}
S. Capozziello, Int. J. Mod. Phys. D 11, 483 (2002), gr-qc/0201033.
S. Capozziello, S. Carloni, and A. Troisi (2003), Rec. Res. Developments in Astronomy
and Astrophysics, Research Signpost Publisher, astro-ph/0303041.
S. M. Carroll, V. Duvvuri, M. Trodden and M. S. Turner, Phys. Rev. D 70, 043528 (2004),
arXiv:astro-ph/0306438.
A.~de la Cruz-Dombriz and A.~Dobado,   
  Phys.\ Rev.\ D {\bf 74}, 087501 (2006)   [gr-qc/0607118];
  P.~K.~S.~Dunsby, E.~Elizalde, R.~Goswami, S.~Odintsov and D.~Saez-Gomez,  
  Phys.\ Rev.\ D {\bf 82}, 023519 (2010)  [arXiv:1005.2205 [gr-qc]];
  S.~Carloni, R.~Goswami and P.~K.~S.~Dunsby,   
  Class.\ Quant.\ Grav.\  {\bf 29}, 135012 (2012)  [arXiv:1005.1840 [gr-qc]];
 E.~Elizalde and D.~Saez-Gomez,   
  Phys.\ Rev.\ D {\bf 80}, 044030 (2009)   [arXiv:0903.2732 [hep-th]];
N.~Goheer, J.~Larena and P.~K.~S.~Dunsby,  
  Phys.\ Rev.\ D {\bf 80}, 061301 (2009)   [arXiv:0906.3860 [gr-qc]];
K. Bamba, S. Capozziello, S. Nojiri and S.~D.~Odintsov,
Astrophys. and Space Science, 342, 155 (2012) 
 arXiv:1205.3421;
  S.~Das, N.~Banerjee and N.~Dadhich,
  Class.\ Quant.\ Grav.\  {\bf 23}, 4159 (2006)
  doi:10.1088/0264-9381/23/12/012
  [astro-ph/0505096];
  G.~J.~Olmo and D.~Rubiera-Garcia,
  Phys.\ Rev.\ D {\bf 84}, 124059 (2011)
  doi:10.1103/PhysRevD.84.124059
  [arXiv:1110.0850 [gr-qc]];
  C.~Bejarano, G.~J.~Olmo and D.~Rubiera-Garcia,
  Phys.\ Rev.\ D {\bf 95}, no. 6, 064043 (2017)
  doi:10.1103/PhysRevD.95.064043
  [arXiv:1702.01292 [hep-th]];
  D.~Bazeia, L.~Losano, R.~Menezes, G.~J.~Olmo and D.~Rubiera-Garcia,
  Eur.\ Phys.\ J.\ C {\bf 75}, no. 12, 569 (2015)
  doi:10.1140/epjc/s10052-015-3803-0
  [arXiv:1411.0897 [hep-th]].
\bibitem{unifying}
  S.~Nojiri and S.~D.~Odintsov,   
  Phys.\ Rev.\ D {\bf 77}, 026007 (2008)
  [arXiv:0710.1738 [hep-th]];
S.Nojiri and S.~D.~Odintsov,
Phys.\ Rev.\  {\bf D68}, 123512 (2003) [hep-th/0307288];
  S.~Nojiri, S.~D.~Odintsov and D.~Saez-Gomez,
  Phys.\ Lett.\ B {\bf 681}, 74 (2009)
  doi:10.1016/j.physletb.2009.09.045
  [arXiv:0908.1269 [hep-th]];
  G.~Cognola, E.~Elizalde, S.~D.~Odintsov, P.~Tretyakov and S.~Zerbini,
  Phys.\ Rev.\ D {\bf 79}, 044001 (2009)
  doi:10.1103/PhysRevD.79.044001
  [arXiv:0810.4989 [gr-qc]].
  
  \bibitem{Joyce:2014kja} 
  A.~Joyce, B.~Jain, J.~Khoury and M.~Trodden,
  Phys.\ Rept.\  {\bf 568}, 1 (2015)
  doi:10.1016/j.physrep.2014.12.002
  [arXiv:1407.0059 [astro-ph.CO]].

\bibitem{Planck-Inflation}   
  P.~A.~R.~Ade {\it et al.} [Planck Collaboration],   
  Astron.\ Astrophys.\  {\bf 571} (2014) A22   [arXiv:1303.5082 [astro-ph.CO]].
  P.~A.~R.~Ade {\it et al.} [BICEP2 and Keck Array Collaborations],
  Phys.\ Rev.\ Lett.\  {\bf 116}, 031302 (2016)
  doi:10.1103/PhysRevLett.116.031302
  [arXiv:1510.09217 [astro-ph.CO]].

\bibitem{Starobinsky1980}
A. A. Starobinsky, Phys. Lett. B 91, (1980) 99.

\bibitem{Khoury:2003rn}
  J.~Khoury and A.~Weltman,
  Phys.\ Rev.\ D {\bf 69}, 044026 (2004)
  doi:10.1103/PhysRevD.69.044026
  [astro-ph/0309411].

\bibitem{Pogosian:2007sw}
  L.~Pogosian and A.~Silvestri,
  Phys.\ Rev.\ D {\bf 77}, 023503 (2008)
  Erratum: [Phys.\ Rev.\ D {\bf 81}, 049901 (2010)]
  doi:10.1103/PhysRevD.77.023503, 10.1103/PhysRevD.81.049901
  [arXiv:0709.0296 [astro-ph]].

\bibitem{Hu:2007nk}
  W.~Hu and I.~Sawicki,
  Phys.\ Rev.\ D {\bf 76}, 064004 (2007)
  doi:10.1103/PhysRevD.76.064004
  [arXiv:0705.1158 [astro-ph]].

  \bibitem{Nojiri:2007cq}
  S.~Nojiri and S.~D.~Odintsov,
  Phys.\ Rev.\ D {\bf 77}, 026007 (2008)
  doi:10.1103/PhysRevD.77.026007
  [arXiv:0710.1738 [hep-th]].

  \bibitem{Appleby:2007vb}
  S.~A.~Appleby and R.~A.~Battye,
  Phys.\ Lett.\ B {\bf 654}, 7 (2007)
  doi:10.1016/j.physletb.2007.08.037
  [arXiv:0705.3199 [astro-ph]];
%
%
  D.~Saez-Gomez,
  Class.\ Quant.\ Grav.\  {\bf 30}, 095008 (2013)
  doi:10.1088/0264-9381/30/9/095008
  [arXiv:1207.5472 [gr-qc]];

  A.~de la Cruz-Dombriz, P.~K.~S.~Dunsby, S.~Kandhai and D.~Saez-Gomez,
  Phys.\ Rev.\ D {\bf 93}, no. 8, 084016 (2016)
  doi:10.1103/PhysRevD.93.084016
  [arXiv:1511.00102 [gr-qc]].


\bibitem{Linder2009}
E. V. Linder, 
 Phys. Rev. D 80 (2009) 123528, arXiv:0905.2962.

\bibitem{nostriexp}
  G.~Cognola, E.~Elizalde, S.~Nojiri, S.~D.~Odintsov, L.~Sebastiani and S.~Zerbini,
  Phys.\ Rev.\ D {\bf 77}, 046009 (2008)
  [arXiv:0712.4017 [hep-th]];
E.~Elizalde, S.~Nojiri, S.~D.~Odintsov, L.~Sebastiani and S.~Zerbini,
  Phys.\ Rev.\ D {\bf 83}, 086006 (2011)
  [arXiv:1012.2280 [hep-th]].
K. Bamba, C. Q. Geng and C. C. Lee, 
JCAP 
 1008 (2010) 021, arXiv:1005.4574.



\bibitem{YangLeeLG2010}
L. Yang, C. C. Lee, L. W. Luo and C. Q. Geng, 
 Phys. Rev. D 82 (2010) 103515 [arXiv:1010.2058].
Y. Chen, C.-Q. Geng, C.-C. Lee, L.-W. Luo and Z.-H. Zhu,  Phys. Rev. D 91 (2015) 044019,
arXiv:1407.4303.

\bibitem{OdintsovSGS:2017}
 S.~D. Odintsov, D. Saez-Chillon Gomez, G.~S. Sharov.
 Eur.\ Phys.\ J.\ C { 77} (2017) 862, arXiv:1709.06800.

\bibitem{Nojiri:2003ni}
   S.~Nojiri and S.~D.~Odintsov,
   Gen.\ Rel.\ Grav.\  {\bf 36} (2004) 1765
   doi:10.1023/B:GERG.0000035950.40718.48
   [hep-th/0308176].
\bibitem{Cognola:2005de}
   G.~Cognola, E.~Elizalde, S.~Nojiri, S.~D.~Odintsov and S.~Zerbini,
   JCAP {\bf 0502} (2005) 010
   doi:10.1088/1475-7516/2005/02/010
   [hep-th/0501096].

\bibitem{Elizalde:2017mrn}
   E.~Elizalde, S.~D.~Odintsov, L.~Sebastiani and R.~Myrzakulov,
late-time acceleration,''
   Nucl.\ Phys.\ B {\bf 921} (2017) 411
   doi:10.1016/j.nuclphysb.2017.06.003
   [arXiv:1706.01879 [gr-qc]].

\bibitem{Myrzakulov:2014hca}
   R.~Myrzakulov, S.~Odintsov and L.~Sebastiani,
   Phys.\ Rev.\ D {\bf 91} (2015) no.8,  083529
   doi:10.1103/PhysRevD.91.083529
   [arXiv:1412.1073 [gr-qc]].


\bibitem{OdintsovOS:2017log}
S.~D.~Odintsov, V.~K.~Oikonomou and L.~Sebastiani,
  Nucl.\ Phys.\ B {\bf 923}, 608 (2017)
  doi:10.1016/j.nuclphysb.2017.08.018
  [arXiv:1708.08346 [gr-qc]].

  \bibitem{Liu:2018hno}
  L.~H.~Liu, T.~Prokopec and A.~A.~Starobinsky,
  arXiv:1806.05407 [gr-qc].

\bibitem{Buchbinder:1992rb}
  I.~L.~Buchbinder, S.~D.~Odintsov and I.~L.~Shapiro,
  ``Effective action in quantum gravity,''
  Bristol, UK: IOP (1992) 413 p.


\bibitem{Union21SNe}
N. Suzuki  et al., 
Astrophys. J. 746 (2012) 85, arXiv:1105.3470; http://supernova.lbl.gov/Union/.


\bibitem{Pantheon17}
D. M. Scolnic  et al., 
 Astrophys. J. 859 (2018) 101, arXiv:1710.00845.


\bibitem{Eisen05}
D. J. Eisenstein { et al.}, 
Astrophys. J. 633 (2005) 560, astro-ph/0501171.


\bibitem{Hdata}
 J. Simon, L. Verde and R. Jimenez, 
{ Phys. Rev.  D} { 71} (2005) 123001, astro-ph/0412269;
 D. Stern,  R. Jimenez,  L. Verde, M. Kamionkowski and S. A. Stanford,
  JCAP { 1002} (2010) 008, arXiv:0907.3149; 
 M. Moresco et al., 
  JCAP {1208} (2012) 006, arXiv:1201.3609;
 C. Zhang { et al.}, 
{ Res. Astron. Astrophys.} { 14} (2014) 1221, arXiv:1207.4541;
 M. Moresco, 
Mon. Not. Roy. Astron. Soc. 450(1) (2015) L16, arXiv:1503.01116;
 M. Moresco et al., 
JCAP  {1605} (2016) 014, arXiv:1601.01701.

\bibitem{Planck13}
Planck Collaboration, P. A. R. Ade { et al.} 
 Astron. Astrophys. 571 (2014) A16, arXiv:1303.5076.
Astron. Astrophys. 594 (2016) A13,
 arXiv:1502.01589 [astro-ph.CO].
%



\bibitem{Sharov16}
 G. S. Sharov,
  JCAP 1606 (2016) 023,  arXiv:1506.05246.

\bibitem{PanSh17}
S. Pan and G. S. Sharov, 
 Mon.\ Not.\ Roy.\ Astron.\ Soc. 472(4) (2017) 4736, 
arXiv:1609.02287.

\bibitem{SharovBPNC17}
 G. S. Sharov, S. Bhattacharya, S. Pan, R. C. Nunes and S. Chakraborty,
Mon.\ Not.\ Roy.\ Astron.\ Soc.\  466(3)  (2017) 3497,  arXiv:1701.00780.

\bibitem{BAOdata}
 W. J. Percival { et al.},
{ Mon. Not. Roy. Astron. Soc.} {401(4)} (2010) 2148, arXiv:0907.1660; 
 E. A. Kazin { et al.}, 
{  Astrophys. J.} {710} (2010) 1444, arXiv:0908.2598; 
 F. Beutler { et al.}, 
 { Mon. Not. Roy. Astron. Soc.} { 416(4)} (2011) 3017, arXiv:1106.3366; 
 C. Blake { et al.}, 
 { Mon. Not. Roy. Astron. Soc.}  { 418(3)} (2011) 1707, arXiv:1108.2635; 
 N. Padmanabhan { et al.}, 
 { Mon. Not. Roy. Astron. Soc.}  { 427(3)} (2012) 2132, arXiv:1202.0090; 
 C-H. Chuang and Y. Wang, 
 { Mon. Not. Roy. Astron. Soc.} { 435(1)} (2013) 255, arXiv:1209.0210; 
 C-H. Chuang { et al.}, 
 { Mon. Not. Roy. Astron. Soc.} { 433(4)} (2013) 3559, arXiv:1303.4486; 
 A. J. Ross { et al.}, 
{ Mon. Not. Roy. Astron. Soc.} { 449(1)} (2015) 835,  arXiv:1409.3242; 
 L. Anderson { et al.},  
 { Mon. Not. Roy. Astron. Soc.} { 441(1)} (2014) 24, arXiv:1312.4877; 
 A. Oka { et al.}, 
 { Mon. Not. Roy. Astron. Soc.} { 439(3)} (2014) 2515, arXiv:1310.2820;  
 A. Font-Ribera { et al.}, 
 JCAP { 1405} (2014) 027, arXiv:1311.1767; 
 T. Delubac { et al.}, 
  Astron. Astrophys. { 574} (2015) A59, arXiv:1404.1801. 

\bibitem{Wang17}
Y. Wang  { et al.}
 Mon. Not. Roy. Astron. Soc. 469(3) (2017) 3762, arXiv:1607.03154.

\bibitem{Ratsimbazafy17}
A. L. Ratsimbazafy et al.
Mon. Not. Roy. Astron. Soc. 467(3) (2017) 3239, arXiv:1702.00418.

\bibitem{WangW2013}
Y. Wang and S. Wang, Phys. Rev. D 88 (2013) 069903, arXiv:1304.4514.

\bibitem{HuangWW2015}
 Q.-G. Huang, K. Wang, S. Wang,  
JCAP 1512 (2015) 022,  arXiv:1509.00969.
%


\bibitem{matterInstability}
   A. D. Dolgov, M. Kawasaki, Phys. Lett. B {\bf 573}  1 (2003) [arXiv:astro-ph/0307285];
M. Soussa and R. Woodard, Gen. Rel. Grav. {\bf 36} 855 (2004).
V. Faraoni, Phys. Rev. D {\bf 74} 104017 (2006) [arXiv:astro-ph/0610734].

\bibitem{CopiDK2004}
C. J. Copi, A. N. Davis, L. M. Krauss,
 Phys. Rev. Lett. 92 (2004) 171301,
arXiv:astro-ph/0311334.

\bibitem{Steigman}
G. Steigman, 
Int. J. Mod. Phys. E 15 (2006) 1, arXiv:astro-ph/0511534;
 V. Simha and G. Steigman,
 JCAP 06 (2008) 16, arXiv:0803.3465.
 
\bibitem{PArthENoPE2018}
R. Consiglio { et al.}
 Comput. Phys. Commun. {\bf 233} (2018) 237, arXiv:1712.04378.





\bibitem{delaCruz-Dombriz:2016bjj}
  A.~de la Cruz-Dombriz, E.~Elizalde, S.~D.~Odintsov and D.~Saez-Gomez,
  JCAP {\bf 1605}, no. 05, 060 (2016)
  doi:10.1088/1475-7516/2016/05/060
  [arXiv:1603.05537 [gr-qc]].

\end{thebibliography}
\end{document}